\documentclass[a4paper,fleqn,usenatbib]{mnras}

\usepackage{graphicx}	

\usepackage{savesym}
\usepackage{amsmath}
\savesymbol{iint}
\usepackage{txfonts}
\restoresymbol{TXF}{iint}

\usepackage[T1]{fontenc}
\usepackage{aecompl}
\title[INOV of NLSy1 galaxies]{Intra-night optical variability characteristics
of different classes of narrow line Seyfert 1 galaxies}

\author[Kshama et al.]{ 
Kshama S. K.$^{1},$\thanks{E-mail: kshama.kurian@iiap.res.in}
Vaidehi S. Paliya$^{2}$, C. S. Stalin$^{1}$
\\
$^{1}$Indian Institute of Astrophysics, Block II, Koramangala, Bangalore, 560034, India\\
$^{2}$Department of Physics and Astronomy, Clemson University, 118 Kinard Laboratory, Clemson, S.C. 29634-0978, USA\\
}

\date{Accepted XXX. Received YYY; in original form ZZZ}

\pubyear{2016}

\begin{document}
\label{firstpage}
\pagerange{\pageref{firstpage}--\pageref{lastpage}}
\maketitle

\begin{abstract}
In a first systematic effort to characterize the intra-night optical
variability (INOV) of different classes of narrow line
Seyfert 1 (NLSy1) galaxies, we have carried out observations on a 
sample of radio-loud (RL) and radio-quiet (RQ)
NLSy1 galaxies. The RL-NLSy1 galaxies are further divided into $\gamma$-ray 
loud (GL) and $\gamma$-ray quiet (GQ) NLSy1 galaxies. Our sample consists of four sets, each set consisting of 
a RQ-NLSy1, a GQ-NLSy1 and a GL-NLSy1 galaxy, closely matched in redshift and optical luminosity. 
Our observations on both RQ and GQ-NLSy1 galaxies consist of a total 
of 19 nights, whereas the  data for
GL-NLSy1 galaxies (18 nights) were taken from literature published 
earlier by us. This enabled us to do  
a comparison of the duty cycle (DC) of different classes of NLSy1 galaxies. 
Using power-enhanced F-test, with a variability threshold of 1\%, we find DCs of about 55\%, 39\% and 0\% for
GL, GQ and RQ-NLSy1 galaxies respectively. 
The  high DC and large 
amplitude of INOV (24.0 $\pm$ 13.7\%) shown by GL-NLSy1 
galaxies relative to the other two classes 
might be due to their inner aligned relativistic jets
having large bulk Lorentz factors. The null DC of RQ-NLSy1 
galaxies could 
mean the presence of low power and/or largely misaligned jets in them.  
However, dividing RL-NLSy1 galaxies into 
low and high optical polarization sources, we find that sources with large 
polarization show somewhat higher DCs (69\%) and amplitudes (29\%) compared to those
with low polarization. This points to a possible link between INOV and 
optical polarization. 
\end{abstract}




\section{Introduction}
One of the defining characteristics of active galactic nuclei (AGN) is that 
they show intensity variations over the complete accessible electromagnetic
spectrum from low energy radio waves to high energy $\gamma$-rays
\citep{1995ARA&A..33..163W}.
Such variations are known to occur on time scales ranging from minutes to 
hours to days \citep{1989Natur.337..627M,1996A&A...305...42H,2010ApJ...722..520A}. Variability, particularly those observed at short time scales of 
the order
of minutes, can be an efficient probe of AGN structure as well as the physical
processes happening close to their central engine not accessible to 
conventional imaging techniques. Among AGN, blazars (that comprise both
flat spectrum radio quasars (FSRQs) and BL Lacertae (BL Lac) objects) whose 
jets are pointed 
close to our line of sight \citep{1995PASP..107..803U} are known to show violent (large amplitude
and short time scale) variations. Such short time scale variations in 
blazars are known across the electromagnetic spectrum including 
$\gamma$-rays \citep{2016ApJ...824L..20A} and X-rays \citep{2015ApJ...811..143P} 
as well as optical \citep{2004MNRAS.348..176S}.

Apart from blazars, other classes of AGN too are known to exhibit rapid 
intensity fluctuations in the optical, hereafter termed as intra-night optical
variability (INOV). Example includes the reports of INOV in radio-quiet quasars
\citep{2003ApJ...586L..25G}, lobe dominated quasars
\citep{2004MNRAS.350..175S} and the weak line quasars 
\citep{2013MNRAS.430.1302G}. Thus, it is now
convincingly established that different classes of quasars show INOV and among
them blazars show large amplitude and high duty cycle (DC) of INOV compared to 
radio-quiet quasars that show low amplitude and low DC of INOV
\citep{2004JApA...25....1S}. The detection of INOV in blazars is usually explained by invoking shocks
propagating down their relativistic jets  \citep{1985ApJ...298..114M}. 
Alternative  models of intrinsic flux variations in blazars  include  instabilities or fluctuations in relativistic jet flows  \citep{1992ApJ...396..469H,1992vob..conf...85M}. 
The clear existence of 
INOV (albeit with low amplitude and low DC) in radio-quiet quasars, had led to 
the idea of the presence
of micro-arcsec scale optical synchrotron jets in them
\citep{2003ApJ...586L..25G}. Hints for the presence of weak radio 
jets in radio-quiet quasars are known from deep VLA observations 
\citep{2001ApJ...562L...5B,1993MNRAS.263..425M}
as well as from the 
presence of hard X-ray tail in some radio-quiet quasars
\citep{2000ApJ...531...52G}.

Though the presence of INOV in luminous quasars is well established
\citep{1989Natur.337..627M,2004MNRAS.348..176S}, their
presence in the less luminous counterparts, the Seyfert galaxies is 
not very clear. Though Seyfert galaxies are known since the discovery of AGN
, a new class of Seyfert galaxies called Narrow-line Seyfert 1 galaxies (NLSy1)
were discovered about three decades ago 
by \cite{1985ApJ...297..166O}. In the
optical spectrum these source have emission lines from the broad line region
narrower than Seyfert 1 galaxies but broader than Seyfert 2 galaxies 
\citep{2000ApJ...538..581R}  with
FWHM ($H_\beta$) less than 2000 km sec$^{-1}$. They have weak O{\sc III} 
$(O{\sc III}/H_{\beta} <3)$  and 
strong Fe II lines.
In addition to the narrow permitted lines from the broad line 
region (attributed to low black hole mass),  NLSy1 
galaxies have excess soft X-ray 
emission and steeper spectra in X-rays when compared to broad line
Seyfert 1 galaxies. They also 
show high accretion rates, close to the Eddington limit 
\citep{2002ApJ...565...78B}. 
In the radio regime, NLSy1 galaxies are found to be mostly radio-quiet. 
However, \cite{2006AJ....132..531K} found that about $7\%$ of the NLSy1 galaxies are 
radio-loud having the radio loudness parameter $R > 10$ (R is defined as the 
ratio of the radio flux density at 5 GHz to the optical flux density at 
440 nm \citep{1989AJ.....98.1195K} and among them 
about $ 2\%-3\% $ are very radio-loud (R $>$ 100). 
This is much smaller than the radio-loud fraction of 15\% known in the quasar category of AGN \citep{1995PASP..107..803U}. The small fraction can be due to 
'selection effects' as the sample was selected based on 
$FWHM(H\beta) < 2000$ km s$^{-1}$ irrespective of the luminosity of the 
source \citep{2006AJ....132..531K}. While a majority of radio-loud NLSy1 (RL-NLSy1) galaxies
are known to have a flat radio spectrum, some of them are found to be compact 
steep spectrum sources. There are also a few reports of rapid optical 
flux variations in NLSy1 galaxies 
\citep{2000NewAR..44..539M,2004ApJ...609...69K}. 

After the launch of the {\it Fermi} Gamma-ray Space Telescope 
(hereafter {\it Fermi}),
in the year 2008, the  Large Area Telescope (LAT) aboard {\it Fermi}
detected for the first time $\gamma$-ray emission from the RL-NLSy1 galaxy, PMN J0948+0022 \citep{2009ApJ...699..976A}. This gave proof of 
the presence of relativistic jet in PMN J0948+0022. Soon after the detection in
$\gamma$-rays, INOV with an amplitude of variability of 0.5 mag in 
few hours was  reported by \cite{2010ApJ...715L.113L}  
in PMN J0948+0022. It was also found to be variable 
by \cite{2011nlsg.confE..59M}. Rapid variations have also been reported in the infra-red bands 
by \cite{2012ApJ...759L..31J}. Subsequent
to the report on the detection of $\gamma$-ray emission in PMN J0948+0022, few 
more RL-NLSy1 galaxies were also reported to be emitters of
$\gamma$-rays by {\it Fermi}.  During the start of this 
observational program five NLSy1 galaxies
were known to be $\gamma$-ray emitters \citep{2011nlsg.confE....F,2015A&A...575A..13F} with high 
significance having test statistics (TS) $>$ 100 and the $\gamma$-ray flux
in the 0.1 $-$ 100 GeV band larger than 5 $\times$ 10$^{-8}$ 
ph cm$^{-2}$ s$^{-1}$. A 10$\sigma$ detection corresponds to a TS value of
100 \citep{1996ApJ...461..396M}.
Since then more $\gamma$-ray emitting RL-NLSy1 galaxies have been discovered, such as 1FHL J1410.4+7408 \citep{2016arXiv160805429M}, 
FBQS J1644+2619 \citep{2015MNRAS.452..520D}  and 
SDSS J122222.55+041315.7 \citep{2015MNRAS.454L..16Y}.
The RL-NLSy1 galaxies which emit $\gamma$-rays are hereafter referred to as $\gamma$-ray loud NLSy1 (GL-NLSy1) galaxies.
Majority of the GL-NLSy1 galaxies
have been studied for INOV and from the observations as of now
available in the literature, it is clear that their INOV properties are
similar to blazars 
\citep{2010ApJ...715L.113L,2013MNRAS.428.2450P,2014ApJ...789..143P,2016ApJ...819..121P}. 
Strong and variable optical polarization are also 
known in three GL-NLSy1 galaxies 
\citep{2014ApJ...794...93M,2013ApJ...775L..26I,2014PASJ...66..108I}. The broad band
spectral energy distribution of GL-NLSy1 galaxies resemble 
FSRQs \citep{2009ApJ...707L.142A,2013ApJ...768...52P,2014ApJ...789..143P}.

High quality observations exist on the INOV properties of radio-loud and
radio-quiet quasars. These studies present ample indirect evidence of the 
presence of relativistic miniature (parsec scale) jets in radio-quiet quasars. 
If this 
observational evidence is considered to be present in the less luminous 
NLSy1 galaxies, it is natural to expect both RL and radio-quiet NLSy1 (RQ-NLSy1) 
galaxies to show INOV, though the amplitude and the prevalence of 
INOV will be different among them. Incidentally, the high DC and amplitude of 
INOV detected in GL-NLSy1 galaxies reinforces their similarity to blazars
and also argues for the detectability of INOV in RQ-NLSy1 galaxies. However, 
as of now, INOV studies of other classes of NLSy1 galaxies are limited to 
few nights of observations \citep{2000NewAR..44..539M,2004ApJ...609...69K}. 
There is thus a need for systematic INOV 
observations of different classes of NLSy1 galaxies (a) to establish the 
presence of INOV in RQ-NLSy1 galaxies and (b) to compare the INOV of 
RQ-NLSy1 galaxies with their radio-loud counterparts. To achieve the
above stated objectives we  have carried out a programme of systematic
INOV observations of a sample of NLSy1 galaxies. We present here the 
results of that programme.

\section{Observations and Data reduction}

\subsection{Sample set}
Our sample consists of radio-loud and radio-quiet NLSy1 galaxies. The RL-NLSy1 galaxies are further divided into $\gamma$-ray loud and $\gamma$-ray quiet (GQ) NLSy1 galaxies based on the detection of $\gamma$-rays by {\it Fermi}. There are four sets of NLSy1 galaxies. This was driven by the 
number of GL-NLSy1 galaxies known as well as their observability using the 2m Himalayan Chandra Telescope (HCT) at Hanle,
during the time this 
monitoring program was initiated.  Each set consists of  
a GL-NLSy1, a GQ-NLSy1 and a RQ-NLSy1 galaxy. To avoid selection biases due
to differences in luminosity and redshift, objects were selected such that
all sources in a particular set have similar optical brightness and redshift. 
The four GL-NLSy1 galaxies are taken from the papers already published by us
in \cite{2013MNRAS.428.2450P,2016ApJ...819..121P}. The remaining four 
RQ NLSy1 galaxies and GQ-NLSy1 galaxies each were taken from the catalogue of 
NLSy1 galaxies by  \cite{2006ApJS..166..128Z}. Our entire sample thus
consists of a total of 12 sources spanning the redshift range 0.06 
$< z < $ 0.66.

The RQ-NLSy1 galaxies in each set are selected such that they are 
truly radio-quiet and not-detected at 1.4 GHz at 1 mJy level in 
the FIRST\footnote{Faint Images of the Radio Sky at Twenty cm: 
http://sundog.stsci.edu} survey. However, GQ-NLSy1 galaxies are
those that have a significant detection in FIRST. For the
GQ and GL-NLSy1 galaxies we estimated the radio-loudness R parameter  
using the formula given by \cite{2002AJ....124.2364I} 
\begin{equation} 
log R=log\left (\frac{F_r}{F_{opt}}\right )=0.4(m-t)
\end{equation}
where $m$ is the SDSS r-band magnitude and $t$ is the radio AB magnitude 
defined as 
\begin{equation}
t=-2.5log\left (\frac{F_{int}}{3631 Jy}  \right )
\end{equation}
here, $F_{int}$ is the integrated radio flux and 3631 Jy is the zero point. 
It is derived using the following relation by \cite{1974ApJS...27...21O} 
\begin{equation}
m_{AB} = -2.5 log(f_{\nu})-48.60 \\
\end{equation}
here, $f_{\nu}$ is flux density in erg cm$^{-2}$ s$^{-1}$ Hz$^{-1}$.  
The properties of the objects observed for INOV are given in Table~\ref{tab:eg_table1}. 

\begin{table*}
	\centering
	\caption{Details of the NLSy1 galaxies that are monitored in this program. Column information are as follows: 
(1)IAU Name (2) Other name (3)right ascension (4) declination (5) R-band magnitude (6) redshift (7) type (8) integrated flux from FIRST (here ND refers to 'Not detected' in FIRST survey) (9) radio loudness parameter and (10) optical polarization}
	\label{tab:eg_table1}
	\begin{tabular}{llllllllll} 
		\hline
		Name & Other Name & $\alpha_{2000}$ & $\delta_{2000}$ & R      & $z$ & Type  & $F_{int}$ & log R & P$_{opt}$\\
		     &            &                 &                 &  (mag) &     &       &  (mJy)         &       & (\%)      \\
		(1) & (2) & (3) & (4) & (5) & (6) & (7) & (8) & (9) & (10)\\
		\hline
		Set-1\\
		\hline
		J0324+3410 & 1H 0323+342                & 03:24:41.10 & +34:10:46.0 & 15.7 & 0.061 & GL-NLSy1 & 559$^*$   & 2.468  & 3$^a$ \\
		J2219+1207 & II Zw 177                  & 22:19:18.53 & +12:07:52.2 & 16.4 & 0.080 & GQ-NLSy1 & 1.51      & 0.196  & -    \\
		J0351-0526 & 2MASX J03510759-0526370    & 03:51:07.61 & -05:26:37.1 & 16.1 & 0.069 & RQ-NLSy1 & ND        & -      & -    \\
		\hline
		Set-2\\
		\hline
		J0849+5108 & SBS 0846+513               & 08:49:58.00 & +51:08:29.0 & 18.3 & 0.586 & GL-NLSy1 & 350.53    & 3.3   & 10$^b$ \\
		J1613+5247 & SDSS J161301.86+524749.5   & 16:13:01.88 & +52:47:49.5 & 18.7 & 0.530 & GQ-NLSy1 & 1.35      & 1.04  & -      \\
		J2123+0102 & SDSS J212309.51+010213.0   & 21:23:09.51 & +01:02:13.1 & 19.2 & 0.590 & RQ-NLSy1 & ND        & -     & -     \\
		\hline
		Set-3\\
		\hline
		J0948+0022 & PMN J0948+0022             & 09:48:57.30 & +00:22:26.0 & 18.4 & 0.585 & GL-NLSy1 & 111.46    & 2.892 & 36$^c$ \\
		J2339-0912 &  FBQS J2339-0912           & 23:39:03.82 & -09:12:21.2 & 19.3 & 0.660 & GQ-NLSy1 & 4.39      & 1.788 &  -    \\
		J1326+0334 &  SDSS J132654.80+033456.3  & 13:26:54.81 & +03:34:56.3 & 18.7 & 0.588 & RQ-NLSy1 & ND        & -     &  -    \\
		\hline 
		Set-4\\
		\hline
		J1505+0326 & PKS 1502+036               & 15:05:06.50 & +03:26:31.0 & 18.2 & 0.409 & GL-NLSy1 & 380.49     & 3.312 & - \\
		J1256+3852 &  FBQS J1256+3852           & 12:56:02.10 & +38:52:31.0 & 17.4 & 0.419 & GQ-NLSy1 & 2.1        & 0.744 & -      \\
		J0037-0933 &  SDSS J003753.72-093331.8  & 00:37:53.73 & -09:33:31.8 & 18.9 & 0.399 & RQ-NLSy1 & ND         & -     & -       \\
		\hline
	\end{tabular}

\footnotesize{
$\ast$ value taken from  \citep{2010A&A...518A..10V}, 
$^a$ \citep{2014PASJ...66..108I}, 
$^b$ \citep{2014ApJ...794...93M},
$^c$ \citep{2013ApJ...775L..26I} 
}

\end{table*}


\subsection{Observations}
The observations of the RQ and GQ-NLSy1 galaxies in our sample were carried out over
a period of 2 years from  2012 August to 2014 September, using the 2 m HCT operated by the Indian Institute of 
Astrophysics, Bangalore. It is a Ritchey-Chretien system with f/9 Cassegrain 
focus. The CCD detector has a gain of 1.22e$^{-}$ and readout noise of 
4.8e$^{-}/ADU$. The field of view is $10'\times 10'$. Each pixel of the CCD 
corresponds to $0.3''\times 0.3''$ of the sky \footnote{http://www.iiap.res.in/centers/iao}.  
The objects were observed in 
Bessel R band as it gives the best 
S/N ratio. The typical S/N ratio in our observations
is around 150 and the time resolution is of the order of 300 seconds for
most of the objects. 
According to \cite{1990PhDT.......263C}, the probability
of detecting INOV in a source is enhanced if it is observed 
continuously for a duration of about 3-4 hours on any given night. Hence, we 
tried to monitor each source for greater than 3 hours continuously 
whenever possible. Also, each source was suitably pointed to have 
two or
more comparison stars within a magnitude of the target NLSy1 galaxies. The 
log of observations is given in Table~\ref{tab:eg_table2}.

\begin{table}
	\centering
	\caption{Log of observations. Columns information are as follows: (1) IAU name 
(2) date of observation (3) duration of observation and (4) number of data points}
	\label{tab:eg_table2}
	\begin{tabular}{llcl} 
		\hline
		Name & Date & $\Delta$T & N\\
		 & & (hours) & \\
		(1) & (2) & (3) & (4)\\
		\hline 
		J2219+1207 & 29/08/2013 & 3.2 & 30\\
			   & 26/08/2014 & 4.5 & 27\\
			   & 22/09/2014 & 7.6 & 73\\
		\hline
		J0351-0526 & 01/02/2014 & 2.9 & 11\\
			   & 02/02/2014 & 2.7 & 20\\
		\hline
		J1613+5247 & 03/07/2013 & 4.7 & 13\\
			   & 12/07/2013 & 6.1 & 25\\
			   & 31/03/2014 & 6.3 & 26\\
		\hline
		J2123+0102 & 11/08/2012 & 5.3 & 28\\
			   & 03/06/2013 & 2.8 & 13\\
		\hline
		J2339-0912 & 05/10/2013 & 6.4 & 24\\
		\hline
		J1326+0334 & 30/03/2014 & 6.6 & 27\\	   
		\hline
		J1256+3852 & 10/04/2013 & 3.2 & 23\\
			   & 05/01/2014 & 3.4 & 10\\
			   & 01/02/2014 & 2.5 & 9\\
			   & 02/02/2014 & 1.6 & 9\\ 
			   & 01/05/2014 & 2.8 & 20\\
		\hline
		J0037-0933 & 16/10/2012 & 5.6 & 12\\
			   & 06/10/2013 & 6.0 & 21\\
		\hline
	\end{tabular}
\end{table}

\subsection{Data Reduction}
Preliminary processing (bias subtraction and flat fielding) of the observations
as well as photometry were done using the IRAF software \footnote{IRAF
is distributed by the National Optical Astronomy Observatories,
which is operated by the Association of Universities for
Research in Astronomy Inc. under cooperative agreement with the 
National Science Foundation}. Instrumental magnitudes of the target NLSy1 
galaxies and the 
comparison stars were determined by aperture photometry using the
{\it phot} task in APPHOT package in IRAF.  At least three comparison 
stars (usually more) with similar brightness to the NLSy1 galaxies were 
considered in each frame. A crucial parameter to the {\it phot} task
is the radius of the aperture for photometry and this 
determines the S/N ratio of any object in the observed frames. According
to \cite{1989PASP..101..616H} the S/N ratio of a 
source in a CCD frame is maximum when the radius of the photometric aperture 
is approximately equal to the full width at half maxima (FWHM) of the point spread 
function (PSF) of the objects and decreases for both larger and smaller apertures. 
Hence, to select the optimum aperture that maximizes the S/N, for each night, a 
range of aperture radii was considered and the aperture that minimizes 
the standard deviation of the differential light curve (DLC) of the 
steadiest pair of comparison stars was taken as the optimum aperture for 
photometry of the target NLSy1 galaxies for that night 
\citep{2004JApA...25....1S}. The standard deviation of the DLC of the steadiest pair of comparison stars thus constructed
will provide the actual error in the photometry, as the photometric
errors reported by IRAF are an underestimate \citep{2013JApA...34..273G}. 
The positions and
the apparent B and R magnitudes (the uncertainties of which can be
up to 0.25 mag) taken from the USNO-B2 
\footnote{http://www.nofs.navy.mil/data/fchpix/} catalogue, of the 
comparison stars for each of the sources, are mentioned in 
Table~\ref{tab:eg_table3}.

\begin{table}
	\centering
	\caption{Properties of the comparison stars used in differential photometry. Column information are as follows: (1) IAU name 
(2) Star identification(S)  (3) right ascension (4) declination (5) B and  (6) R magnitudes}
	\label{tab:eg_table3}
	\begin{tabular}{llcccc} 
		\hline
		IAU Name  & S & $\alpha_{2000}$  & $\delta_{2000}$  & B & R\\
		 & & & & (mag) & (mag)\\
		(1) & (2) & (3) & (4) & (5) & (6)\\
		\hline 
		J2219+1207 & S1 & 22:19:26.09 & +12:05:17.10 & 16.27 & 14.17\\
			   & S2 & 22:19:30.62 &	+12:04:34.68 & 16.10 & 14.46\\
			   & S3 & 22:19:21.36 &	+12:10:07.77 & 16.62 & 14.76\\
		\hline
		J0351-0526 & S1 & 03:51:00.82 & -05:26:04.03 & 16.15 & 14.21\\
			   & S2 & 03:51:13.20 & -05:23:19.85 & 16.13 & 13.93\\
			   & S3 & 03:50:57.94 & -05:27:45.46 & 17.14 & 14.96\\
		\hline
		J1613+5247 & S1 & 16:12:42.85 & +52:46:30.54 & 18.10 & 17.8\\
			   & S2 & 16:13:05.83 & +52:44:33.85 & 20.88 & 18.87\\
			   & S3 & 16:13:00.55 &	+52:45:19.09 & 20.89 & 19.28\\
		\hline
		J2123+0102 & S1 & 21:23:03.32 & +01:02:56.14 & 18.56 & 16.59\\ 
			   & S2 & 21:23:19.18 & +01:04:14.39 & 17.88 & 16.15\\
			   & S3 & 21:23:13.89 & +01:07:52.62 & 20.76 & 18.53\\
		\hline
		J2339-0912 & S1 & 23:38:52.98 & -09:13:19.34 & 20.22 & 17.84\\
			   & S2 & 23:39:02.42 & -09:11:45.71 & 18.57 & 17.88\\
			   & S3 & 23:39:09.57 & -09:08:33.28 & 17.52 & 16.94\\
		\hline
		J1326+0334 & S1 & 13:26:45.36 & +03:33:30.48 & 20.64 & 18.78\\   
			   & S2 & 13:26:53.83 &	+03:34:34.47 & 19.57 & 18.08\\
			   & S3 & 13:26:52.20 &	+03:33:04.39 & 18.26 & 17.05\\
		\hline
		J1256+3852 & S1 & 12:55:46.11 & +38:49:00.05 & 17.25 & 16.86\\
			   & S2 & 12:56:14.05 & +38:48:58.61 & 16.80 & 15.94\\ 
			   & S3 & 12:56:16.92 & +38:48:20.20 & 18.44 & 16.77\\
		\hline
		J0037-0933 & S1 & 00:37:40.28 & -09:32:40.10 & 18.84 & 17.89\\
			   & S2 & 00:37:44.14 & -09:33:25.37 & 19.13 & 17.65\\
			   & S3 & 00:37:50.49 & -09:30:50.38 & 19.47 & 18.83\\
		\hline
	\end{tabular}
\end{table}

\section{Analysis}

Using the instrumental magnitudes obtained from photometry, 
DLCs of the NLSy1 galaxies were generated relative
to a minimum of three comparison stars, using an optimum aperture
that is close to the median FWHM on that night. In most of the NLSy1 galaxies
in our sample, the host galaxy has negligible effects on the photometry 
of the objects \citep{2000AJ....119.1534C}, as the central AGN 
dominates. However, for the NLSy1 galaxies in Set 1 (lowest redshift bin), it is likely that the underlying host galaxy may affect 
the flux variations of the NLSy1 galaxy owing to varying seeing conditions. As noted by \cite{1991AJ....101.1196C}, for sources with underlying host galaxy 
components, any spurious variations introduced by fluctuations in atmospheric 
seeing are typically smaller than the observational uncertainties. 
Even so, for unambiguous detection of INOV and to rule out
any spurious variations contributed by the hosts of the NLSy1 galaxies, 
we have checked for any close correspondence between the seeing 
variations in any particular night and the DLCs of the NLSy1 galaxies
relative to the comparison stars. We find that the host galaxies have negligible effect on the photometry reported here. We define INOV as flux variations 
greater than or of the order of 1\% which corresponds to 
95\% confidence (2$\sigma$ level). This is driven by the error in 
our photometry which is typically around 0.005 magnitudes. To check 
for the presence or 
absence of INOV in a source, we have used the power-enhanced F test 
\citep{2014AJ....148...93D}.

\subsection{Power-enhanced F-test}
This test is proposed to be free from the difficulties associated with the
widely used criteria to test INOV, such as the C and F statistics
\citep{2014AJ....148...93D} and has found increased usage in recent 
studies of INOV in AGN \citep{2015MNRAS.452.4263G,2016MNRAS.460.3950P}.
It consists of transforming the comparison star's DLCs to have the same 
photometric noise, as if their magnitudes exactly matched the mean magnitude 
of the AGN under study. The enhanced F, statistical criterion is
defined as  
\begin{equation}
F_{enh} = \frac{s_{qso}^2}{s_c^2}
\end{equation}
Here, $s_{qso}^2$ is the variance of the NLSy1 galaxy-reference star DLC 
and $s_c^2$ is the stacked variance \citep{2014AJ....148...93D} of the comparison star-reference star 
DLCs given as
\begin{equation}
s_c^2=\frac{1}{(\sum_{j=1}^k N_j - k)}\sum_{j=1}^{k}\sum_{i=1}^{N_j}s_{j,i}^2
\end{equation}
where $N_j$ is the number of observations of the $j^{th}$ star, $k$ is 
the total number of comparison stars. $s_{j,i}^2$ is the 
scaled square deviation, defined as
\begin{equation}
s_{j,i}^2=\omega_j(m_{j,i}-\bar{m_j})^2
\end{equation}
where, $\omega_j$ is the scaling factor, $m_{j,i}$'s are the differential instrumental
magnitudes and $\bar{m_j}$ is the mean differential magnitude of the reference star and the $j^{th}$ comparison star.
Following \cite{2011MNRAS.412.2717J}, we have taken 
$\omega_j$ as the ratio of the averaged square error of the differential 
instrumental
magnitudes in the NLSy1 galaxy - reference star DLC 
(<$\sigma_q^2$>) to the  averaged square error of the differential instrumental
magnitudes in the comparison star - reference star 
DLC (<$\sigma_{s_j}^2$>). 
For the $j^{th}$ star DLC
\begin{equation}
\omega_j=\frac{<\sigma_q^2>}{<\sigma_{s_j}^2>}
\end{equation}
It is now known that the photometric errors returned by IRAF are
underestimated \citep{2013JApA...34..273G,2004JApA...25....1S} by factors
of about 1.5. But as $\omega_j$ is the ratio of the errors, this factor gets
cancelled out. 
Thus by stacking the variances of the comparison stars, the degrees of freedom of 
the denominator of the F-distribution given in Equation 4 increases, which 
in turn 
increases the power of the test \citep{2015AJ....150...44D}. In this work, 
for most of the nights we have three stars from which the star with magnitude
as close as possible or brighter than the NLSy1 galaxy in the field
is taken as the reference star. Hence we are left with two stars as the comparison 
stars. The quasar and comparison stars have the same number of observations 
and hence the number of degrees of freedom of the numerator and denominator 
in Equation 4 are $\nu_1=N-1$ and $\nu_2=k(N-1)$ respectively.  
The $F_{enh}$ value is then estimated and compared with the critical value of F 
($F_{c}$) for $\alpha =$ 0.05 which corresponds to 95\% confidence 
level. A source is considered variable if $F_{enh}$ is greater than or equal to
the calculated $F_{c}$. The estimated values of $F_{enh}$ are given in 
Table~\ref{tab:eg_table4}.

\subsection{Amplitude of Variability}
For objects that are found to satisfy the adopted
statistical criterion, we calculated the variability amplitude (Amp) 
following \cite{1996A&A...305...42H}. Amp is defined as 
\begin{equation}
	Amp = \sqrt{(A_{max}-A_{min})^2-2\sigma^2}
\end{equation}
here, $A_{max}$ is the maximum in the NLSy1 galaxy - reference 
star DLC, $A_{min}$ is the minimum in the NLSy1 galaxy - reference 
star DLC and $\sigma$ is the standard deviation of the 
steadiest comparison star-reference star DLC. 
The calculated amplitude of variability for the variable 
NLSy1 galaxies are shown in 
Table~\ref{tab:eg_table4}. Table~\ref{tab:eg_table5} shows the mean variability amplitude and the associated error in amplitude for the different types of NLSy1 galaxies.

\begin{figure}
    \centering
	\includegraphics[width=6cm]{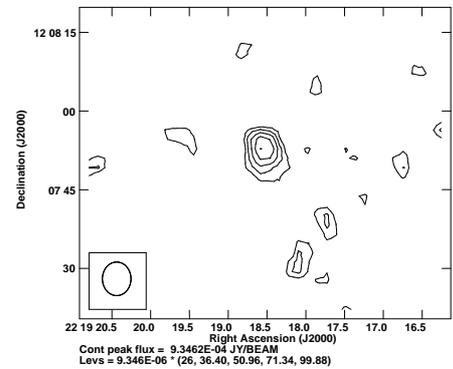}
    \caption{FIRST image of the GQ-NLSy1 galaxy, J2219+1207.}
    \label{fig:fig1}
\end{figure}

\begin{figure*}
     	\centering
	\hbox
	{
	\includegraphics[width=4.5cm,height=8cm]{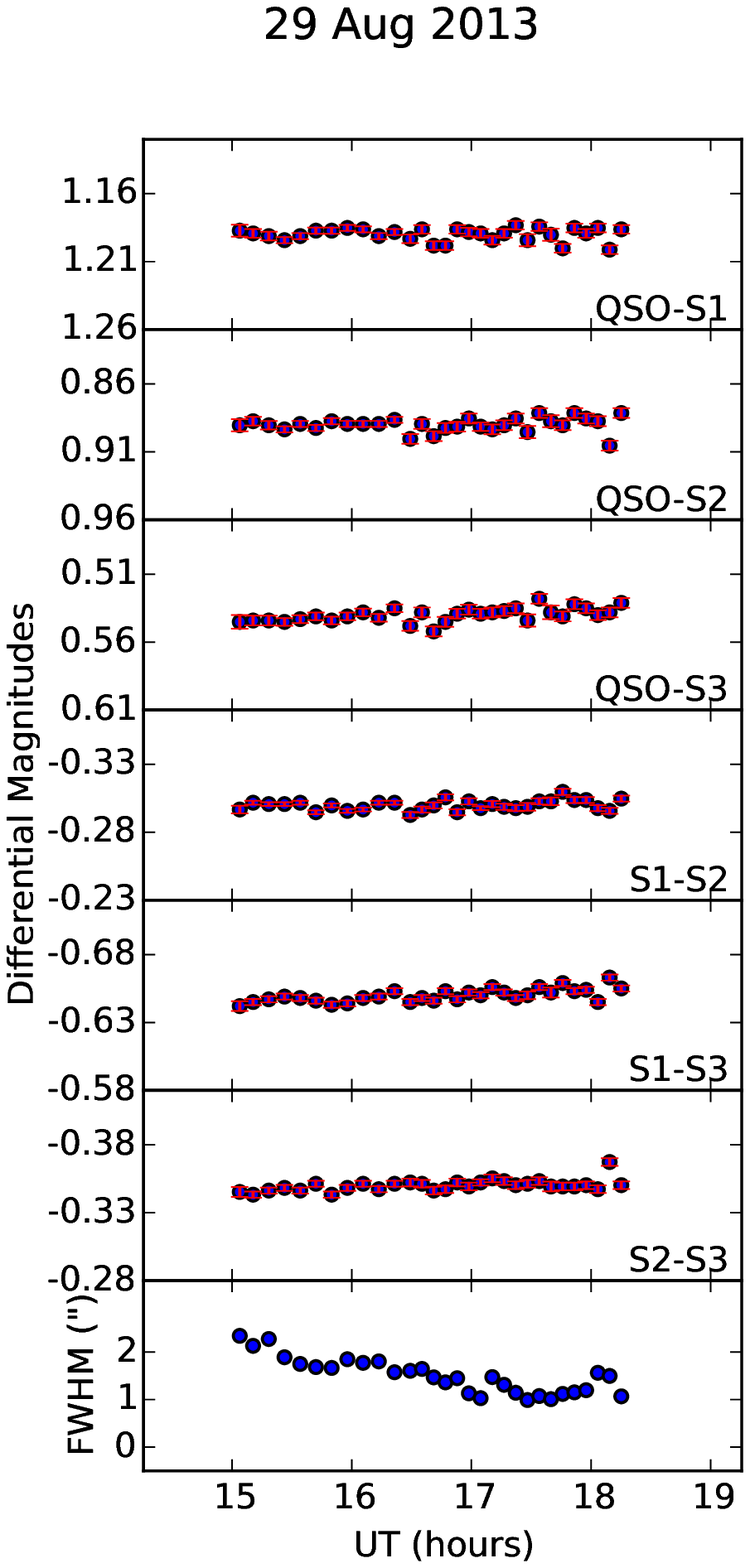}
	\includegraphics[width=4.5cm,height=8cm]{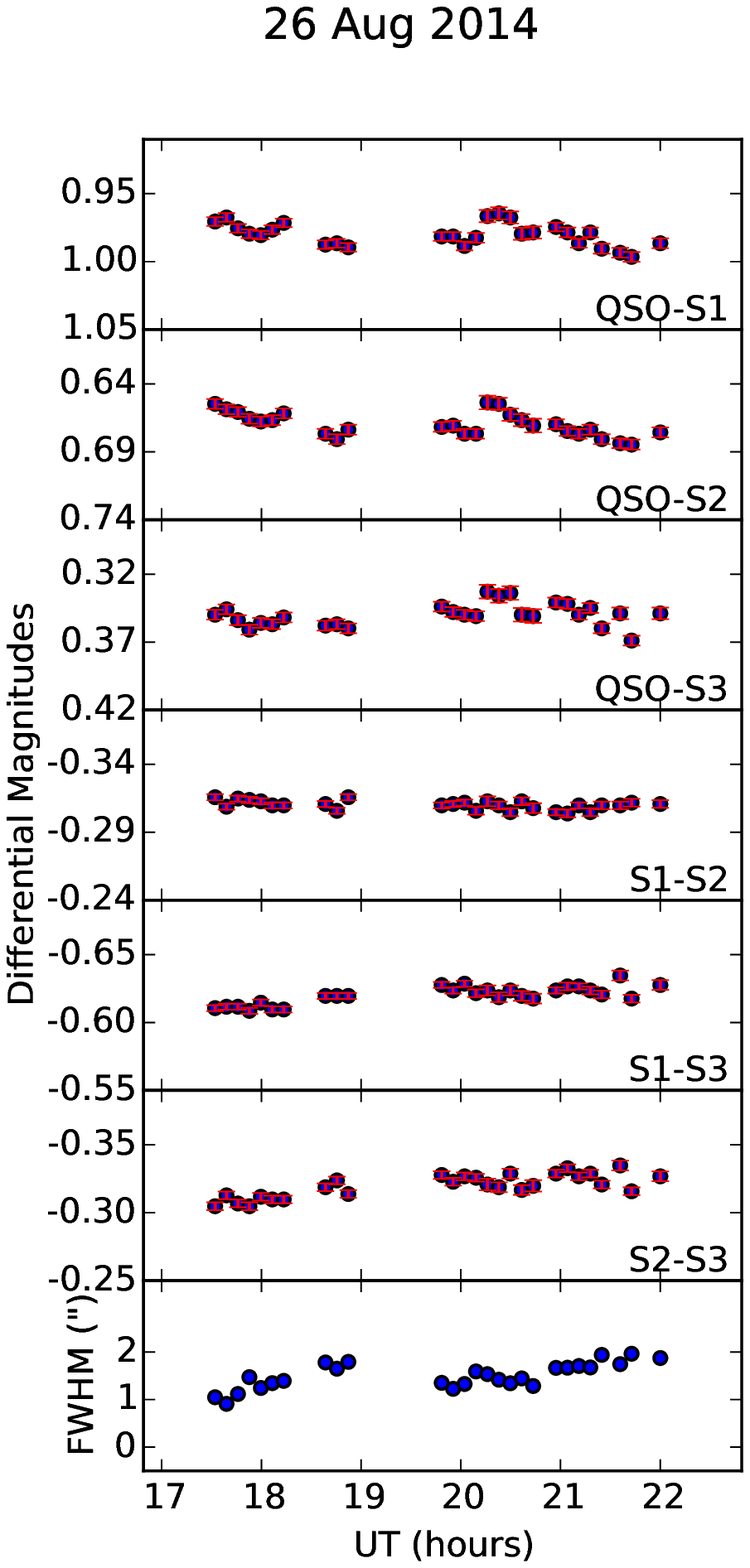}
	\includegraphics[width=6.5cm,height=8cm]{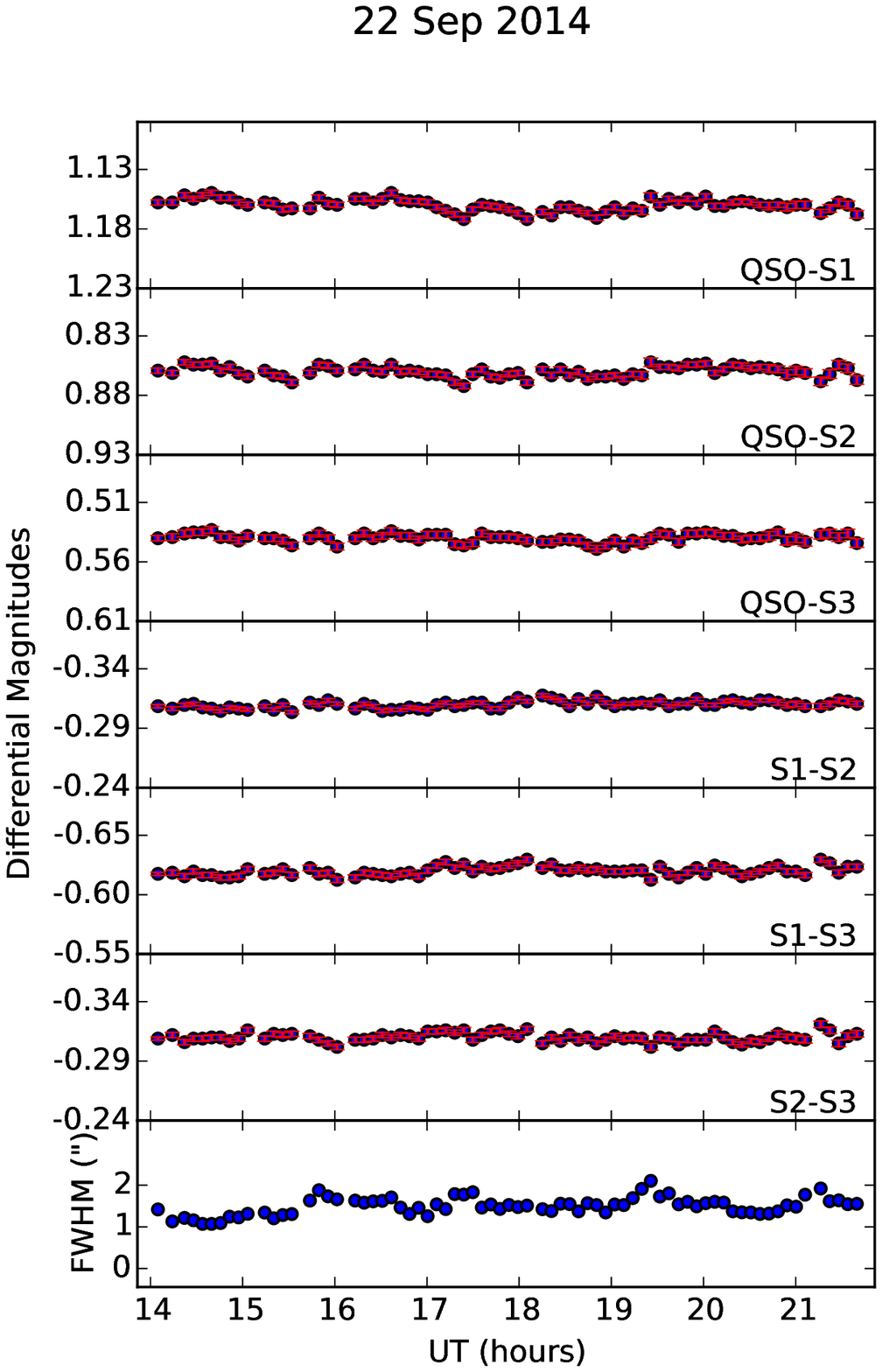}
	}
    \caption{DLCs of the GQ-NLSy1 galaxy, J2219+1207.}
    \label{fig:fig2}
\end{figure*}

\begin{figure}
    \centering
	\hbox
	{
	\includegraphics[width=4.2cm,height=8cm]{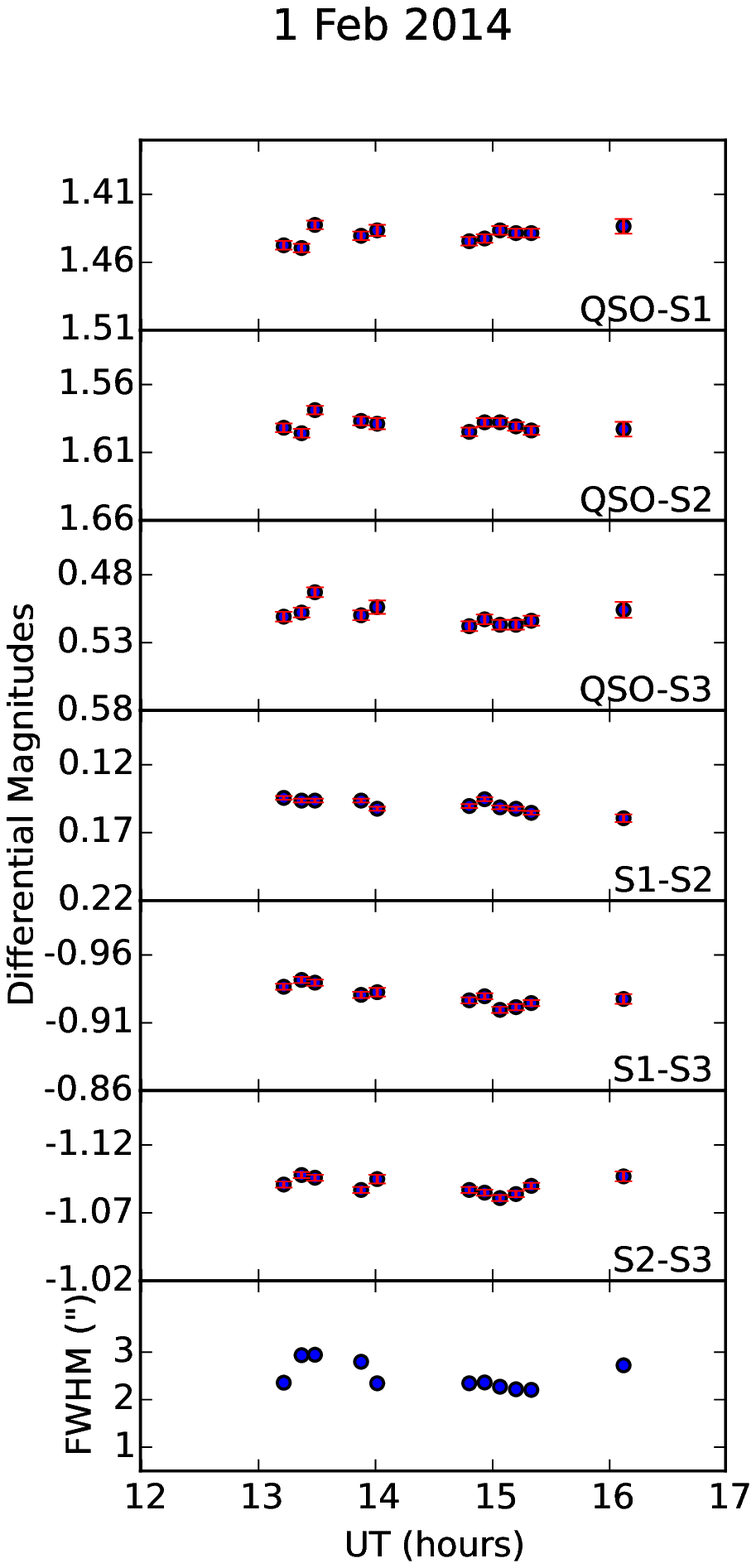}
	\includegraphics[width=4.2cm,height=8cm]{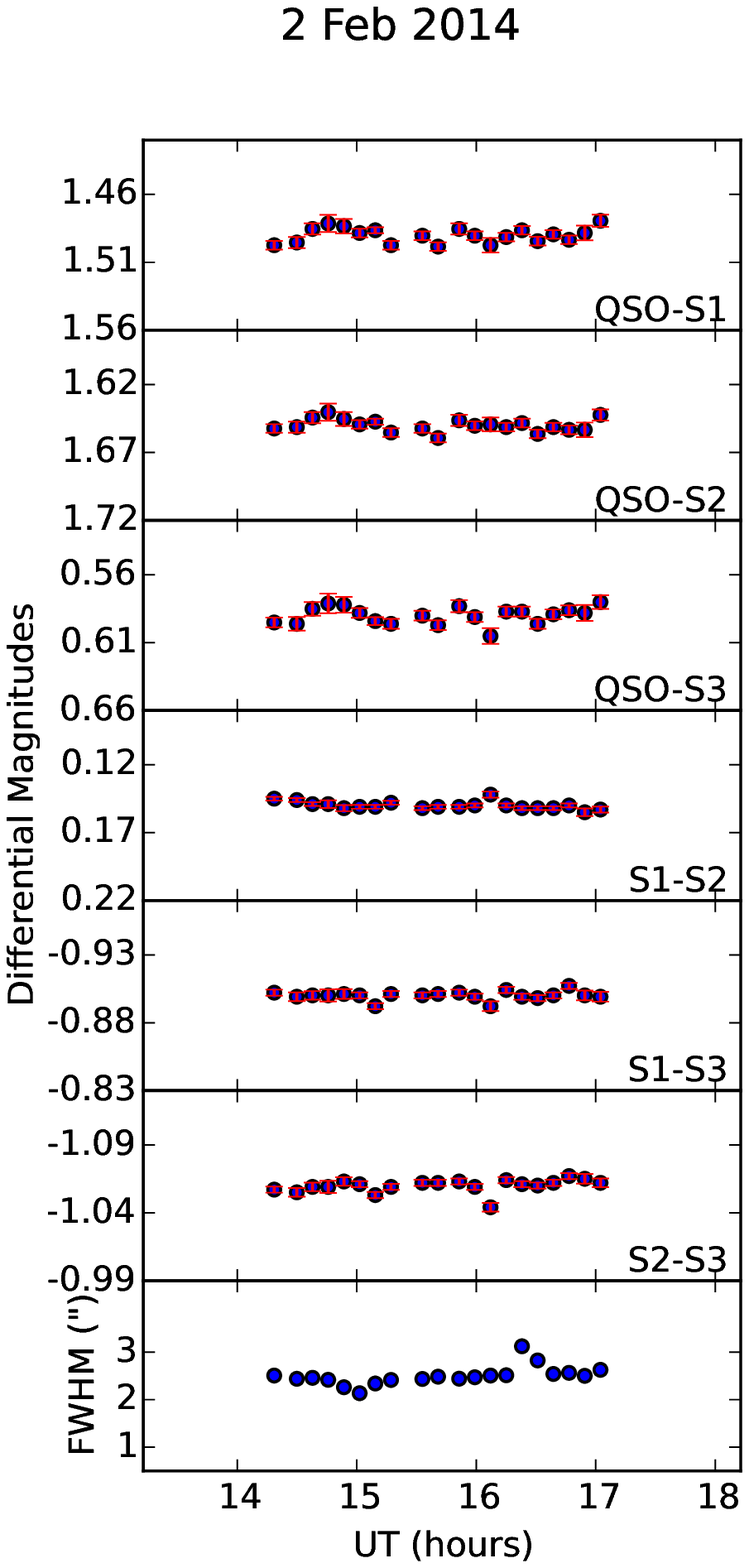}
	}
    \caption{DLCs of the RQ-NLSy1 galaxy, J0351-0526.}
    \label{fig:fig3}
\end{figure}

\begin{figure*}
    \centering
	\hbox
	{
	\includegraphics[width=4.5cm,height=8cm]{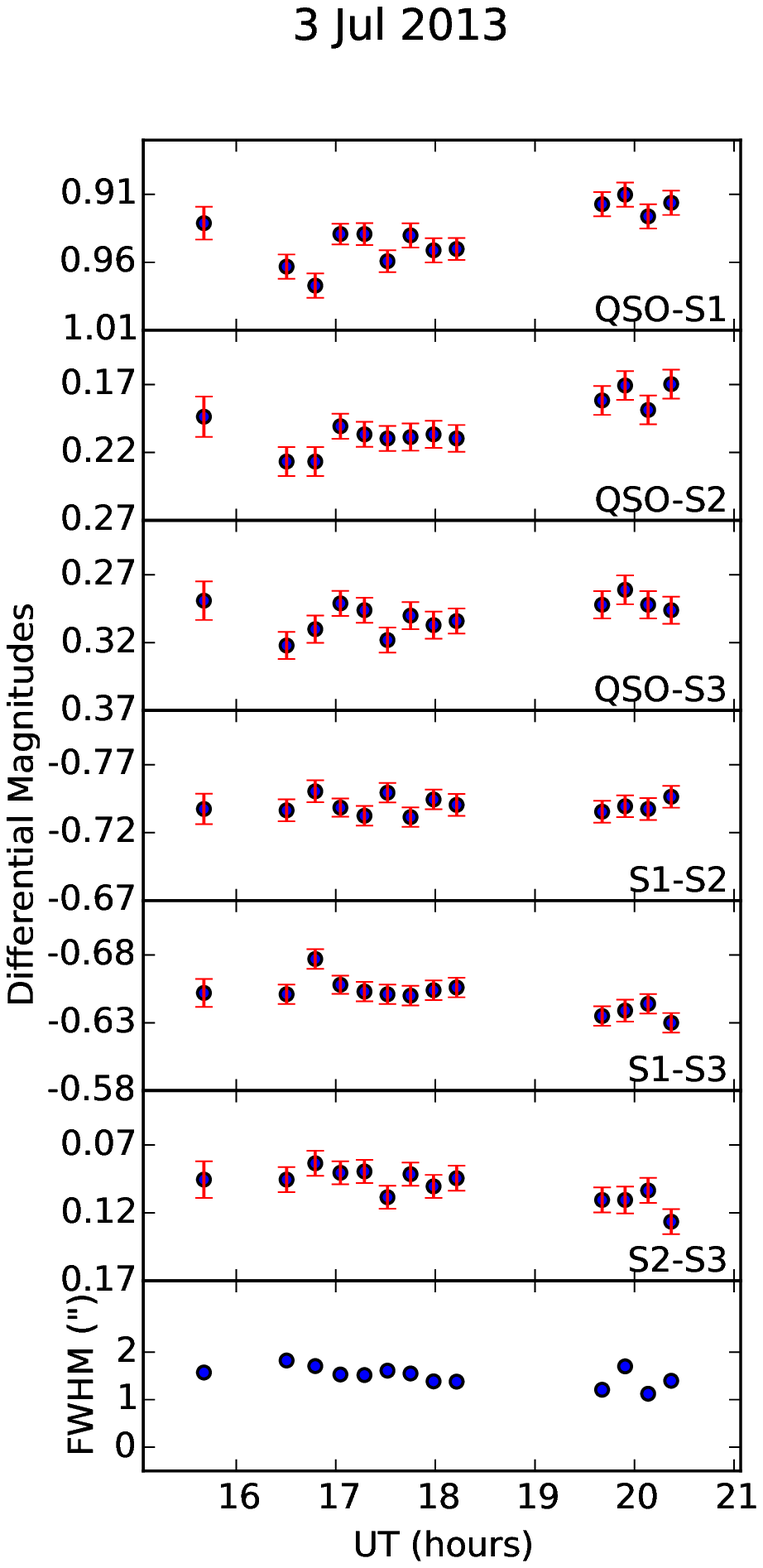}
	\includegraphics[width=6cm,height=8cm]{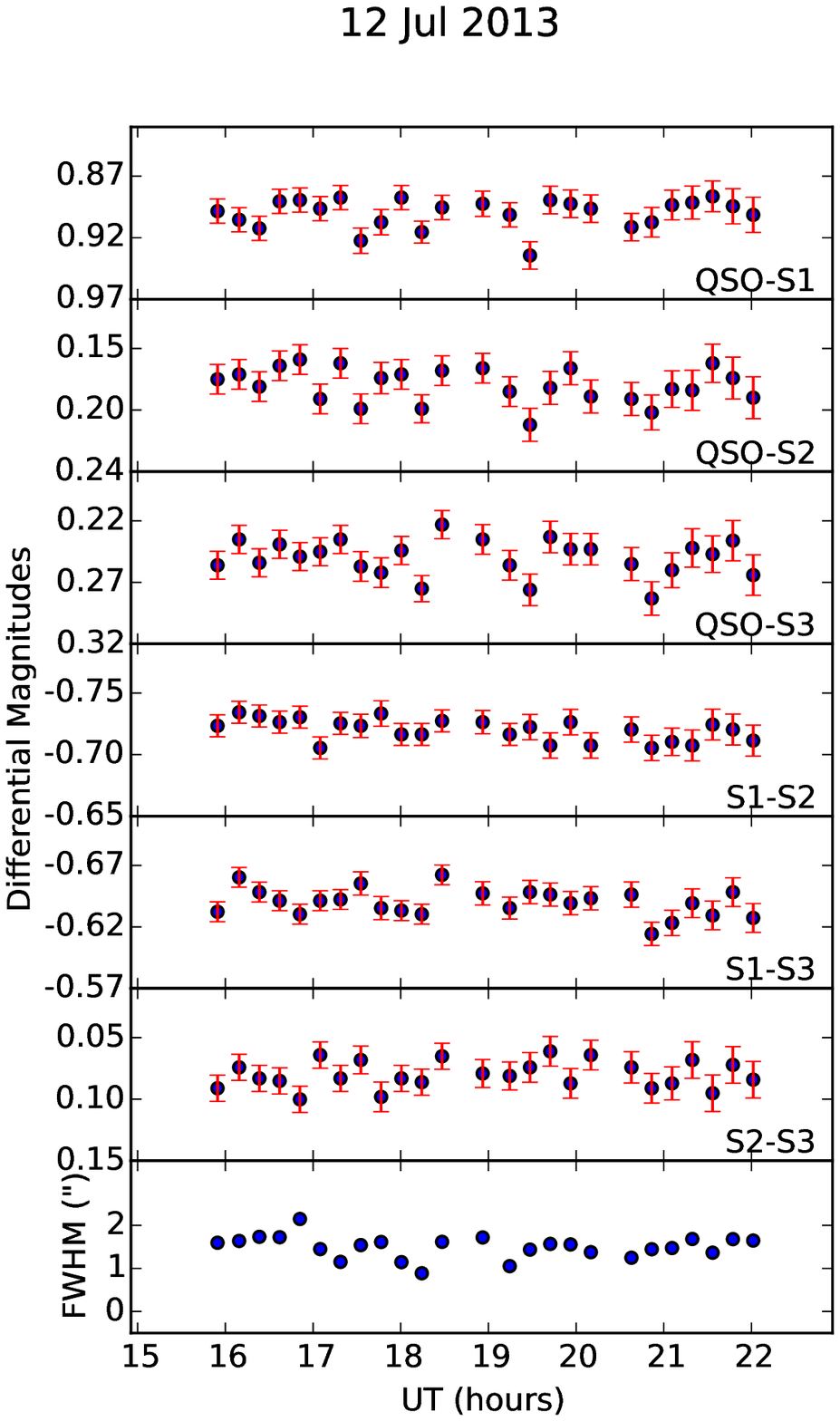}
	\includegraphics[width=6cm,height=8cm]{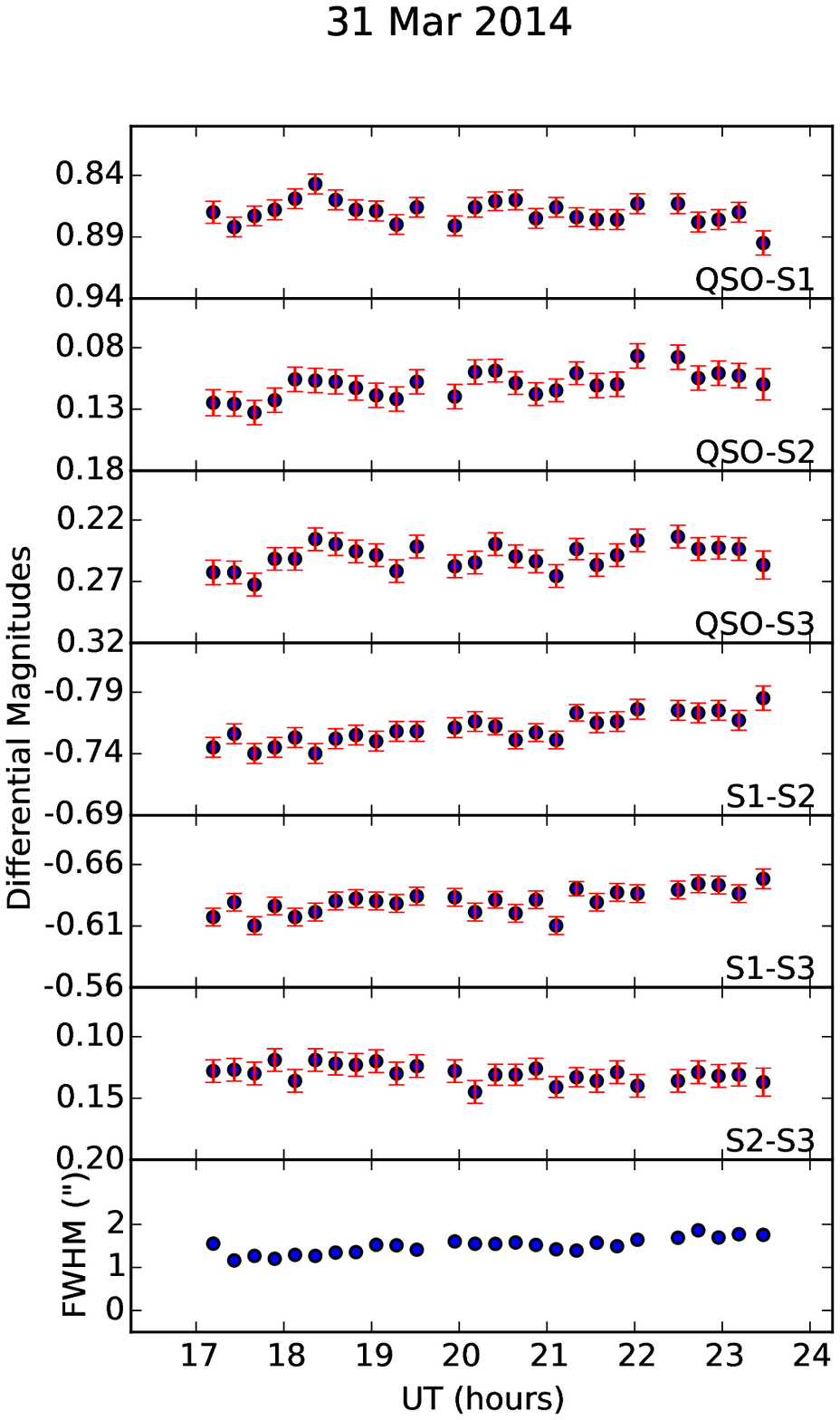}
	}
    \caption{DLCs of the GQ-NLSy1 galaxy, J1613+5247.}
    \label{fig:fig4}
\end{figure*}

\begin{figure}
    \centering
	\hbox
	{
	\includegraphics[width=4.5cm,height=8cm]{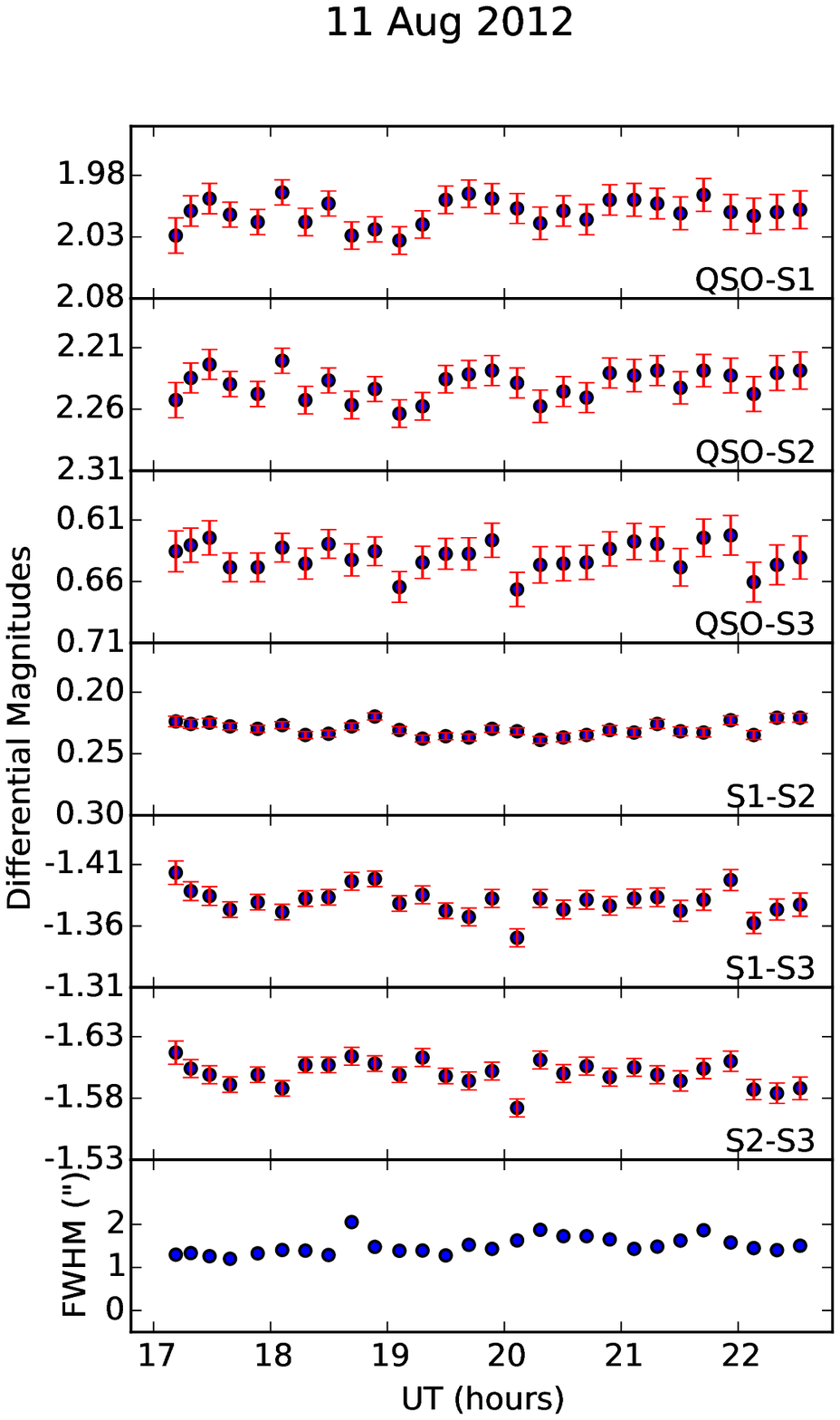}
	\includegraphics[width=4.0cm,height=8cm]{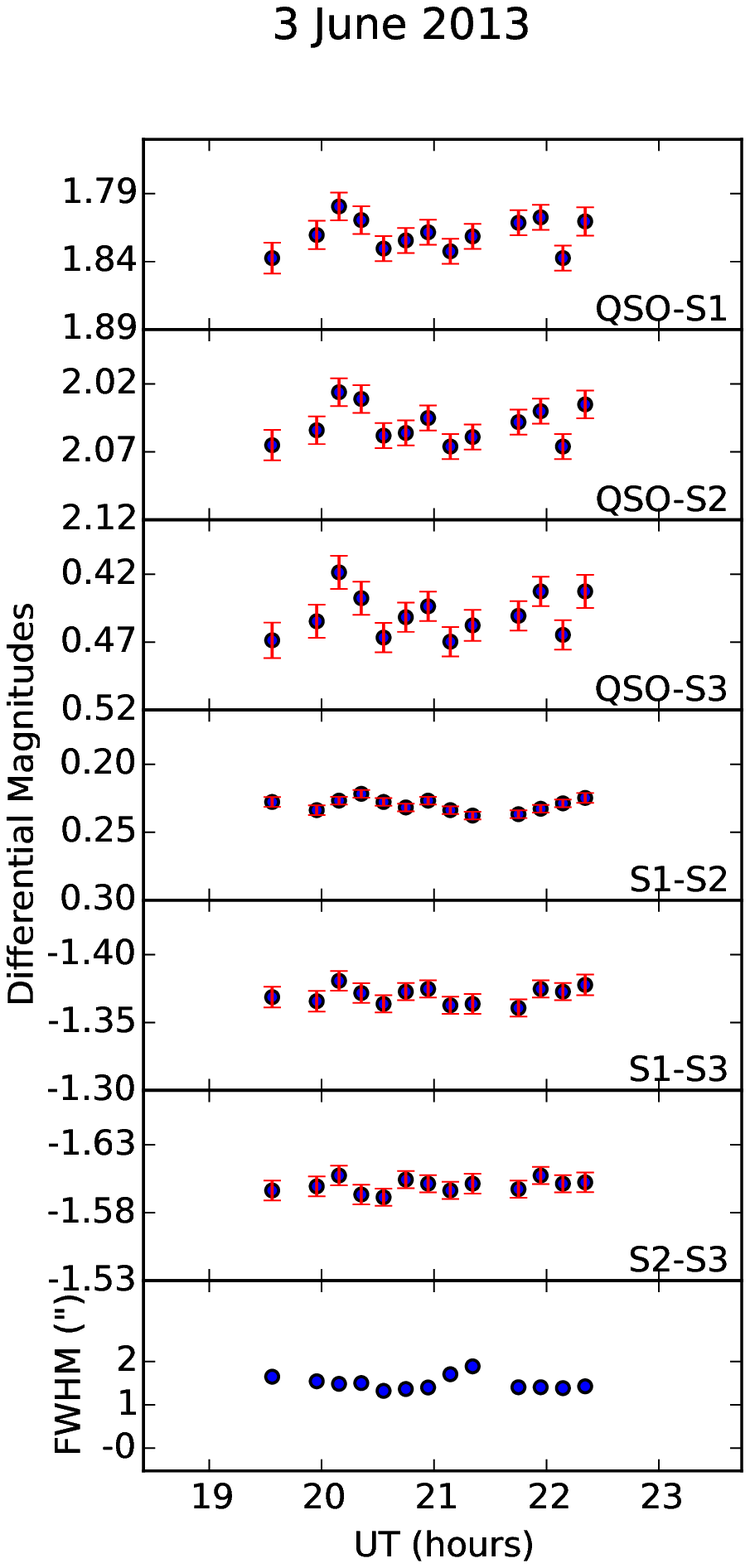}
	}
    \caption{DLCs of the RQ-NLSy1 galaxy, J2123+0102.}
    \label{fig:fig5}
\end{figure}

\begin{figure}
	\centering
	\includegraphics[width=4.5cm, height=8cm]{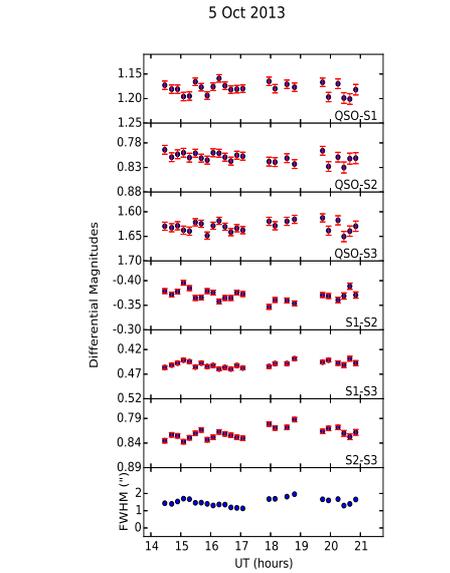}
    \caption{DLC of the GQ-NLSy1 galaxy, J2339-0912.}
    \label{fig:fig6}
\end{figure}

\begin{figure}
	\centering
	\includegraphics[width=4.5cm, height=8cm]{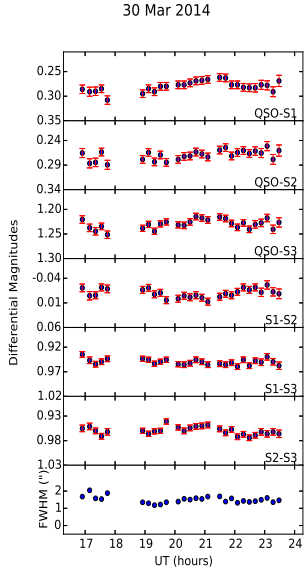}
    \caption{DLC of the RQ-NLSy1 galaxy, J1326+0334.}
    \label{fig:fig7}
\end{figure}

\begin{figure*}
     \centering
	\hbox
	{
	\includegraphics[width=3.6cm,height=8cm]{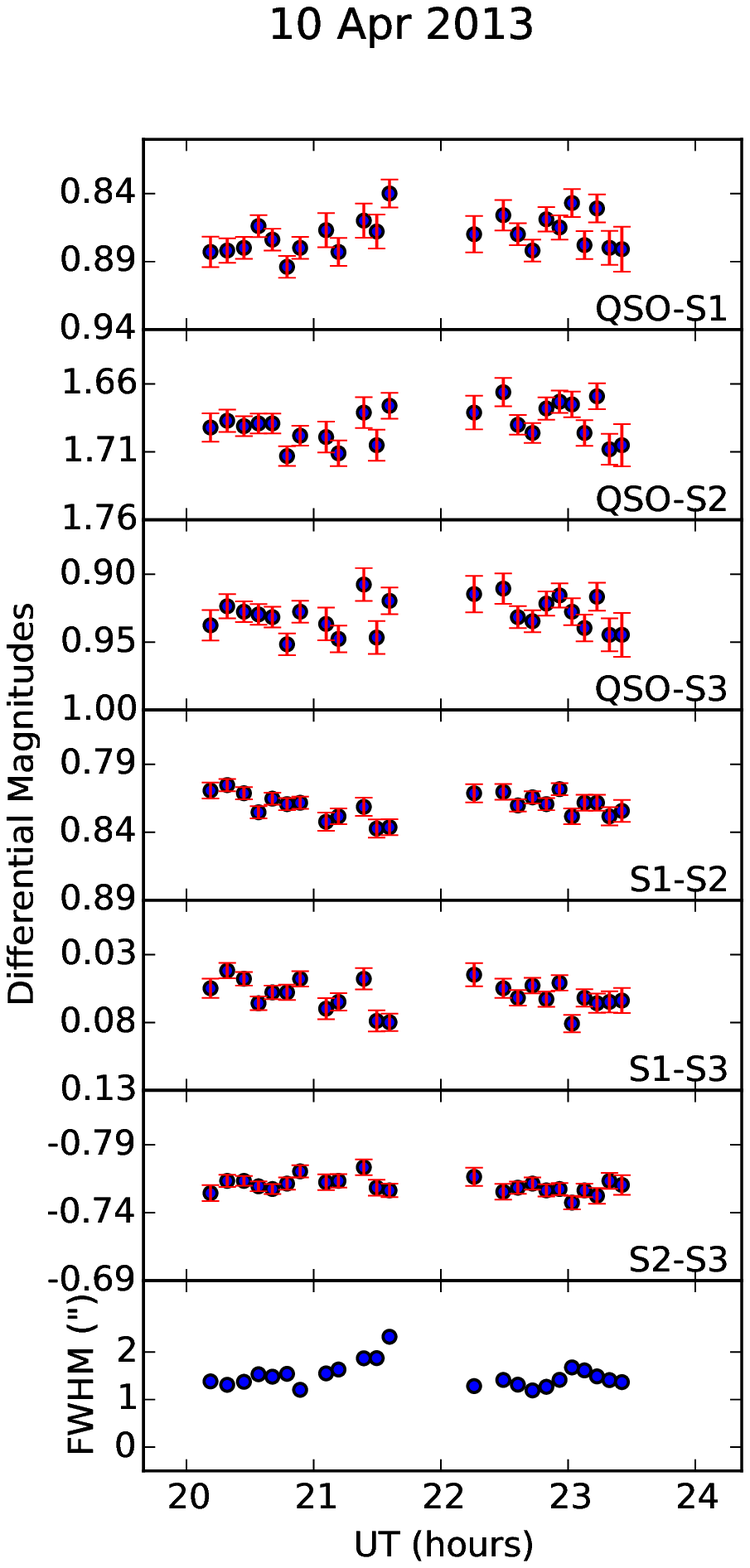}
	\includegraphics[width=3.3cm,height=8cm]{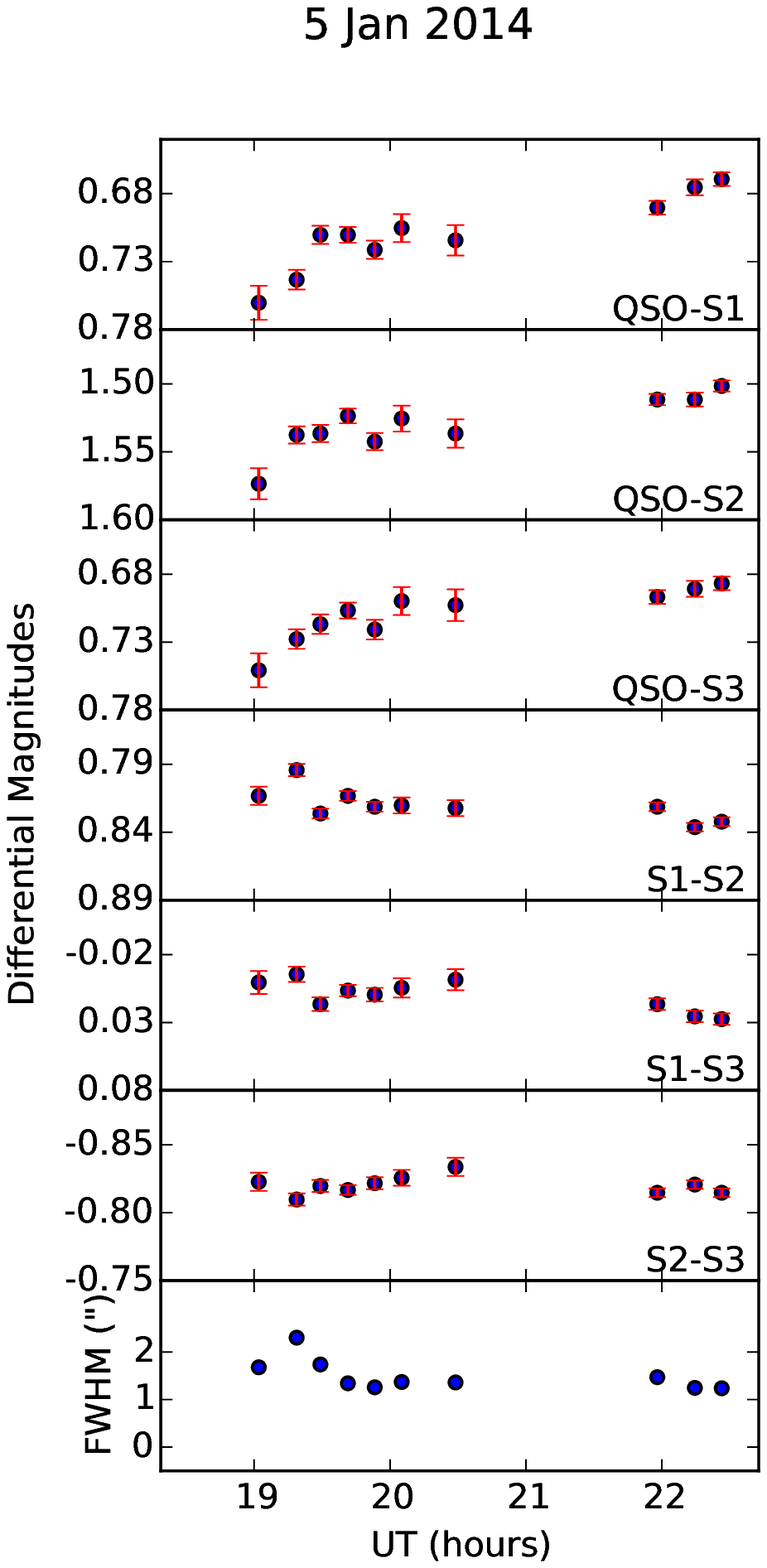}
	\includegraphics[width=3.3cm,height=8cm]{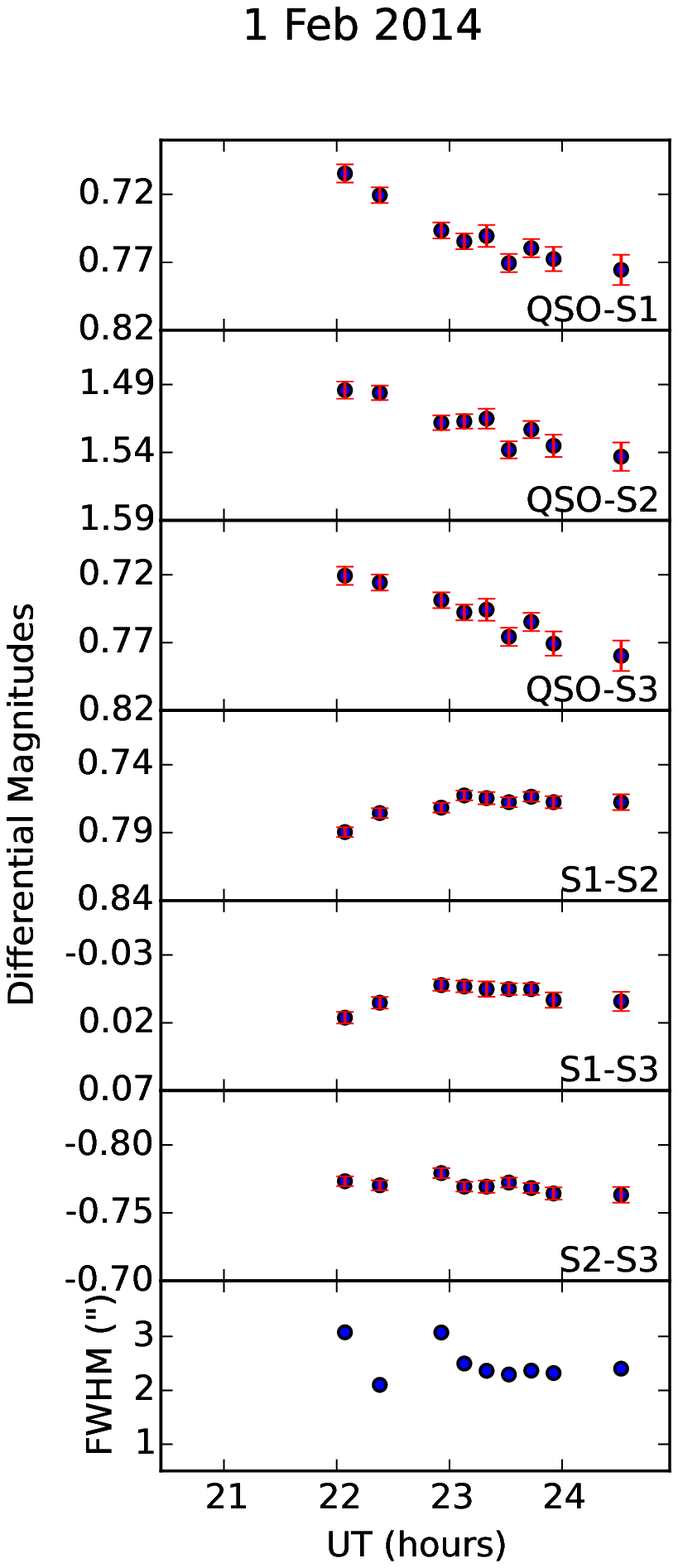}
	\includegraphics[width=3.3cm,height=8cm]{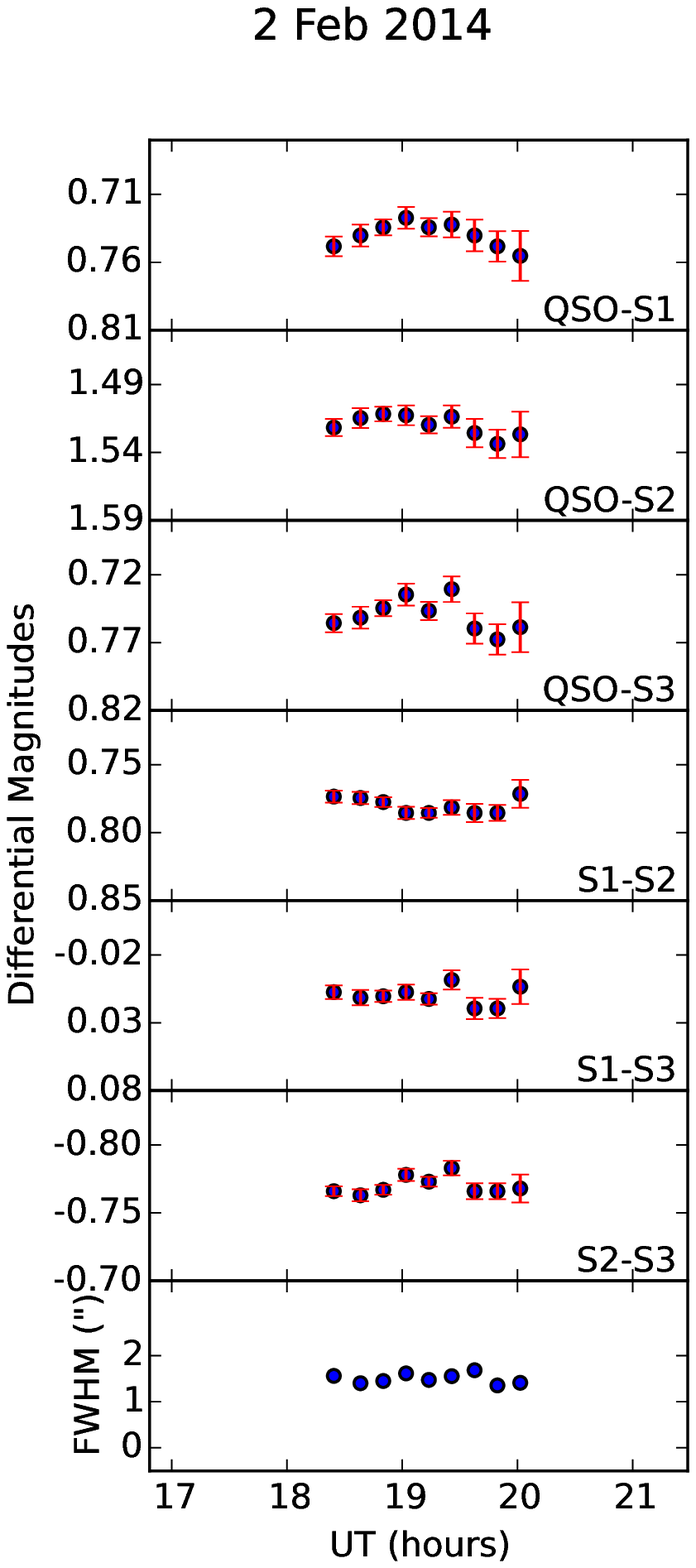}
	\includegraphics[width=3.6cm,height=8cm]{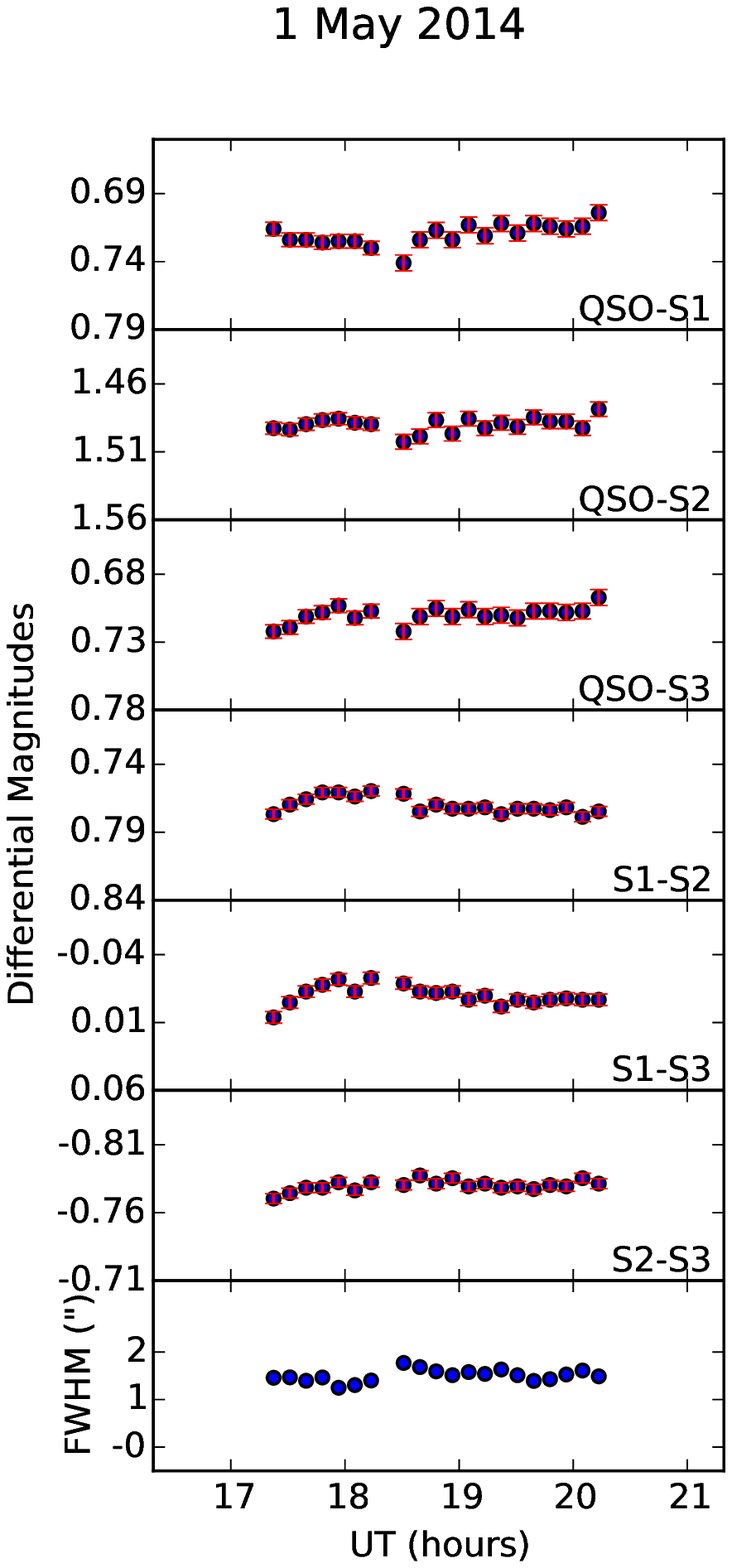}
	}
    \caption{DLCs of the GQ-NLSy1 galaxy, J1256+3852.}
    \label{fig:fig8}
\end{figure*}

\begin{figure}
     \centering
	\hbox
	{
	\includegraphics[width=4.2cm,height=8cm]{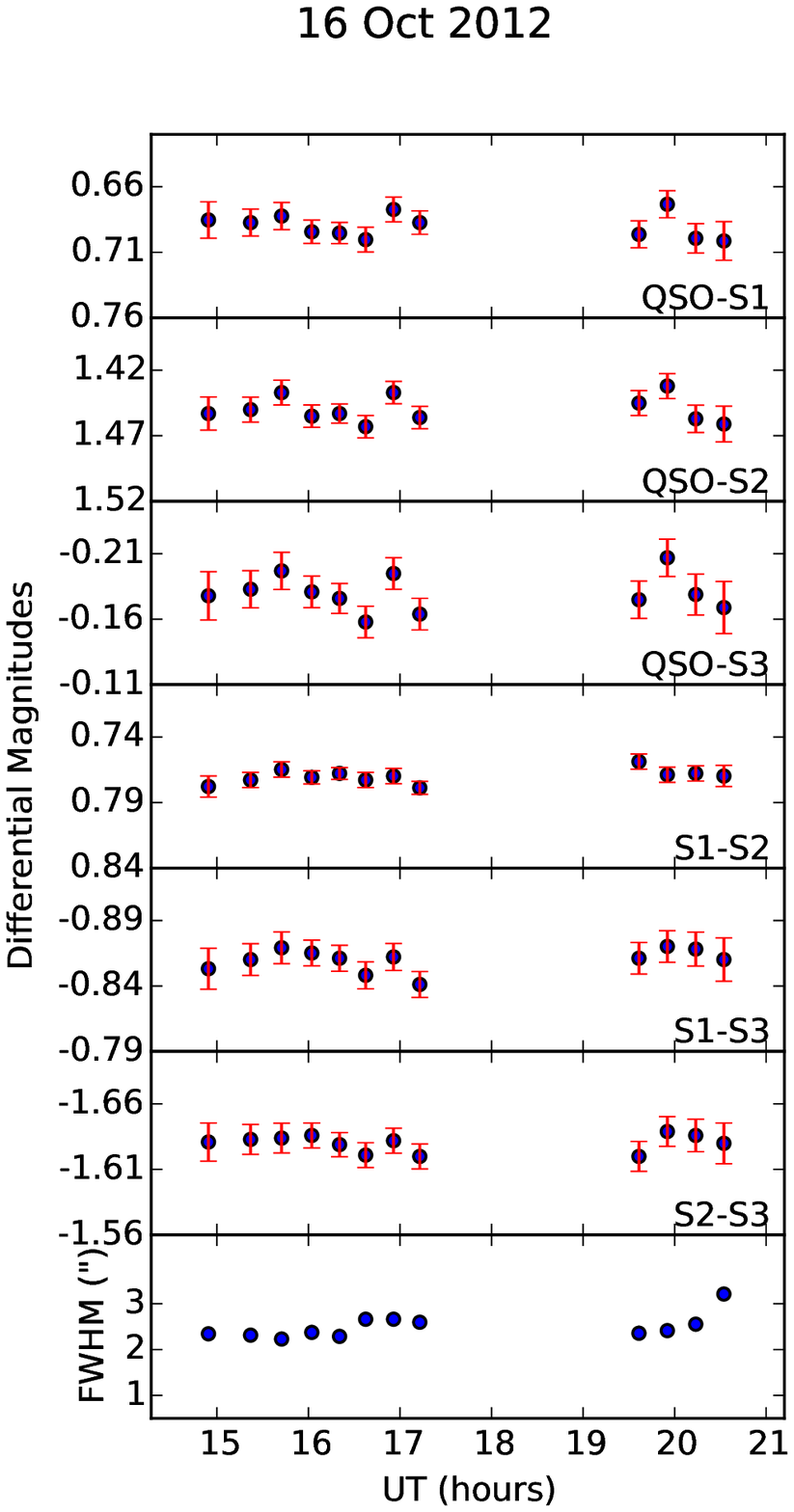}
	\includegraphics[width=4.2cm,height=8cm]{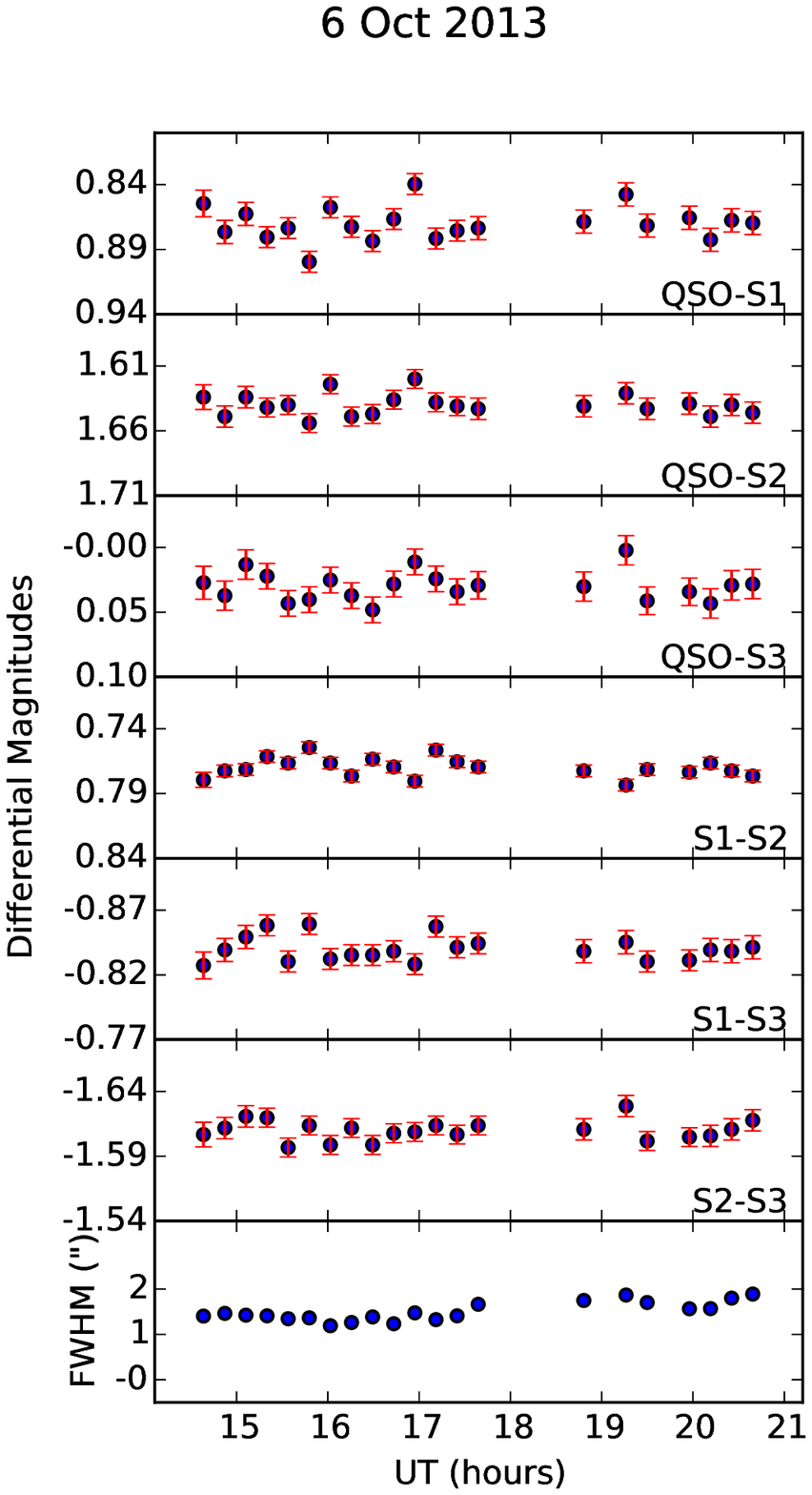}
	}
    \caption{DLCs of the RQ-NLSy1 galaxy, J0037-0933.}
    \label{fig:fig9}
\end{figure}

\begin{table*}
	\centering
	\caption{Results of variability analysis. NV: non-variable and V: variable. Column information are as follows 
(1) IAU name (2) type (3) date of observation (4) reference star (5) comparison stars 
(6) degrees of freedom of numerator ($\nu_1$) and denominator ($\nu_2$) (7) enhanced F test value (8) critical F value for $\alpha =$0.05 (9) enhanced F statistics variability status (10) amplitude of variability}
	\bigskip
	\label{tab:eg_table4}
	\begin{tabular}{cccccccccc} 
		\hline 
		Name & Type & Date & Ref.  & Comparison & $dof$         & $F_{enh}$ & $F_c$(95\%) & Status & Amp \\
		     &      &      & star & stars      & $\nu_1,\nu_2$ &           &       &        & (\%) \\
		(1)  & (2)  & (3)  & (4)  & (5)        & (6)           & (7)       &  (8)  & (9)    & (10) \\
		\hline 
                J0324+3410 & GL & 24/01/12 & S1 & S2,S3 & 104,208 & 1.39 & 1.31 & V & 12.7\\
                           &    & 25/01/12 & S1 & S2,S3 & 46,92   & 0.98 & 1.50 & NV & - \\
                           &    & 26/01/12 & S1 & S2,S3 & 88,176  & 3.96 & 1.34 & V & 7.4\\
                           &    & 02/02/12 & S1 & S2,S3 & 34,68   & 1.05 & 1.60 & NV & - \\
		\hline
		J2219+1207 & GQ & 29/08/13 & S1 & S2,S3 & 29,58 & 0.63 & 1.66 & NV & -\\
			   &    & 26/08/14 & S1 & S2    & 26,26 & 3.36 & 1.93 & V & 3.1\\
			   &    & 22/09/14 & S1 & S2,S3 & 72,144 & 1.33 & 1.39 & NV & -\\
		\hline
		J0351-0526 & RQ & 01/02/14 & S2 & S1,S3 & 10,20 & 0.26 & 2.35 & NV & -\\
			   &    & 02/02/14 & S2 & S1,S3 & 19,38 & 0.47 & 1.87 & NV & -\\
		\hline
		\hline
                J0849+5108 & GL & 20/11/12 & S2 & S1 & 39,39 & 1.97 & 1.70 & V & 12.0\\
                           &    & 09/12/12 & S2 & S1 & 20,20 & 11.87 & 2.12 & V & 24.4 \\
                           &    & 10/12/12 & S2 & S1 & 19,19 & 0.63 & 2.17 & NV & -\\
                           &    & 25/12/12 & S2 & S1 & 44,44 & 1.28 & 1.65 & NV & -\\
                           &    & 12/02/13 & S2 & S1 & 37,37 & 2.38 & 1.73 & V & 36.8\\
                           &    & 11/03/13 & S2 & S1 & 56,56 & 12.01 & 1.56 & V & 33.3\\
		\hline
		J1613+5247 & GQ & 03/07/13 & S1 & S2 & 12,12 & 8.70 & 2.69 & V & 6.6\\
			   &    & 12/07/13 & S1 & S2 & 24,24 & 1.47 & 1.98 & NV & -\\
			   &    & 31/03/14 & S2 & S3 & 25,25 & 2.31 & 1.96 & V & 4.4\\
		\hline
		J2123+0102 & RQ & 11/08/12 & S2 & S1 & 27,27 & 0.32 & 1.90 & NV & -\\
			   &    & 03/06/13 & S2 & S1,S3 & 12,24 & 1.31 & 2.18 & NV & -\\
		\hline
		\hline
                J0948+0022 & GL & 26/01/12 & S1 & S2,S3 & 164,328 & 16.61 & 1.24 & V & 51.9\\
                           &    & 02/02/12 & S1 & S2,S3 & 17,34 & 6.51 & 1.93 & V & 17.1\\
                           &    & 11/03/12 & S1 & S2,S3 & 39,78 & 45.84 & 1.55 & V & 33.1\\
                           &    & 19/04/12 & S1 & S2,S3 & 72,144 & 5.8 & 1.39 & V & 25.2\\
		\hline
		J2339-0912 & GQ & 05/10/13 & S3 & S1,S2 & 23,46 & 0.37 & 1.77 & NV & -\\
		\hline
		J1326+0334 & RQ & 30/03/14 & S3 & S1,S2 & 26,52	& 1.38 & 1.71 & NV & -\\
		\hline
		\hline
                J1505+0326 & GL & 18/04/12 & S1 & S2,S3 & 23,46 & 0.65 & 1.77 & NV & - \\
                           &    & 22/05/12 & S1 & S2,S3 & 10,20 & 1.10 & 2.35 & NV & - \\
                           &    & 23/05/12 & S1 & S2,S3 & 12,24 & 0.51 & 2.18 & NV & - \\
                           &    & 24/05/12 & S1 & S2,S3 & 11,22 & 5.44 & 2.26 & V & 10.1\\
		\hline
		J1256+3852 & GQ & 10/04/13 & S2 & S3 & 22,22 & 1.61 & 2.05 & NV & -\\
			   &    & 05/01/14 & S2 & S3 & 9,9 & 3.91 & 3.18 & V & 7.1\\
			   &    & 01/02/14 & S2 & S3 & 8,8 & 4.49 & 3.44 & V & 4.8\\
			   &    & 02/02/14 & S2 & S1,S3 & 8,16 & 0.54 & 2.59 & NV & -\\
			   &    & 01/05/14 & S2 & S1,S3 & 19,38 & 0.58 & 1.87 & NV & -\\
		\hline
		J0037-0933 & RQ & 16/10/12 & S2 & S1,S3 & 11,22 & 1.86 & 2.26 & NV & -\\
			   &    & 06/10/13 & S2 & S1,S3 & 20,40 & 0.60 & 1.84 & NV & -\\
		\hline
	\end{tabular}
\end{table*}

\subsection{Duty Cycle}
AGN do not show variability at all times they are monitored. Therefore, duty 
cycle (DC) of variability of each class of NLSy1 galaxy 
is estimated by the ratio of the time over which each object in a given
class is found to be variable to the total observing time spent on 
monitoring the objects in a class. As per the definition given by 
\cite{1999A&AS..135..477R}, the DC is calculated as 
\begin{equation}
	DC=100\frac{\sum_{n}^{i=1}N_i(1/\Delta t_i)}{\sum_{n}^{i=1}(1/\Delta t_i)} \%
\end{equation}
where $\Delta t_i=\Delta t_{i,obs}(1+z)^{-1}$ is the duration of monitoring 
session of a source on the i$^{th}$ night, corrected for its cosmological 
redshift $z$. $N_i$ is set to 1 if INOV is detected, else it is set to 0. The results of DC for different classes of 
NLSy1 galaxies are given in Table~\ref{tab:eg_table5}.

\section{Notes on individual objects monitored in this work}

{\bf Set 1 - J2219+1207:} This is a compact radio-loud source 
having an integrated flux density of 1.51 mJy at 1.4GHz. The radio contour of 
the source from FIRST survey generated using AIPS \footnote{AIPS  
is produced and maintained by the National Radio Astronomy Observatory, a 
facility of the National Science Foundation operated under
cooperative agreement by Associated Universities, Inc.} is shown in Fig.~\ref{fig:fig1}. 
It has a radio loudness parameter (log R) smaller than 1 which is the 
dividing value between radio-loud and radio-quiet AGN as  proposed by 
\cite{1989AJ.....98.1195K}. This low value might be attributed to an overestimate
of the optical flux due the underlying host galaxy which is clearly visible in 
the optical images. This source has been observed for three nights over a 
period of 1 year for durations ranging between about 3 to 7.6 hours. 
For statistical calculations, S1 is taken as the reference star and 
S2, S3 are taken as the comparison stars on all the nights
except 26 August 2014 on which only S2 is used as the comparison 
star as S3 varies (Fig.~\ref{fig:fig2}). Seeing is found 
to be steady on all the three nights of observations. On the first night, the 
object is found to be non-variable (NV). However it shows unambiguous 
INOV of the order of 0.03 magnitudes on the second night. On the third
night, when it was monitored for the longest duration, 
the densely sampled light curves with 73 epochs of observation over
7.6 hours shows no INOV. In the long term, it has shown 
both brightening and dimming behaviour during our observations.
Between 29 August 2013 and 26 August 2014 the source has brightened by ~0.2 mag and faded by the same amount when observed a month later on
29 September 2016.

{\bf Set 1 - J0351-0526:} This RQ-NLSy1 galaxy was monitored for two consecutive 
days in February 2014 for about 3 hours on both the nights. For statistical
analysis of variability, the reference star used is S2. S1 and 
S3 are used as the comparison stars (Fig.~\ref{fig:fig3}). It is 
classified as non-variable.

{\bf Set 2 - J1613+5247:} This radio-loud NLSy1 galaxy with log R $=$ 1.04 was
observed for three nights over a period of eight months between
July 2013 and March 2014. The DLC of stars S1 and S2 has the lowest standard
deviation of 0.0076 on two nights and thus has been considered as the steadiest 
pair of stars on those two nights for statistical analysis of variability
of which S1 is taken as the reference star (Fig.~\ref{fig:fig4}).
However, on one night (31 March 2014) stars S2 and S3 are taken as the 
reference star and comparison star respectively, as the star S1 is found to 
be unsteady. This source has shown
definite INOV on 03 July 2013 with INOV amplitude of 6.6\% and on 31 March 2014 with amplitude 4.4\%. On 12 July 2013
it is non-variable.

{\bf Set 2 - J2123+0102:} This radio-quiet source was monitored for two 
nights. The reference star used for the statistical analysis is S2. Stars S1, S3 are used for comparison except on 11 August 2012 where only S1 is used as S3 varies (Fig.~\ref{fig:fig5}). 
The source is non-variable on both the nights.
Long term optical variability is seen in this source, wherein, it
faded by about 0.2 magnitudes over a 10 month period between 11 August 2012 and 3 June 2013.  

{\bf Set 3 - J2339-0912:} This radio-loud NLSy1 galaxy has the highest radio
flux density of 4.39 mJy at 1.4 GHz with a radio-loudness parameter
(log R) of 1.79, close to the dividing line between radio-loud and very radio-loud
NLSy1 galaxies \citep{1989AJ.....98.1195K}. It has a steep radio spectral index
$\alpha$ = $-$1.4 between 6 and 20cm \citep{2007ApJ...667..673G} and the NVSS contour map points to an extended structure. It did not show any unambiguous 
evidence of INOV on the single night it was observed (Fig.~\ref{fig:fig6}).
For statistical tests, the comparison stars used are S1 and S2,
whereas S3 is taken as the reference star.   

{\bf Set 3 - J1326+0334:} This RQ-NLSy1 was observed for only one epoch. 
Based on the statistical test, we classify this source as NV on 
that night (Fig.~\ref{fig:fig7}). The reference star used is S3 and 
S1, S2 are taken as  the comparison stars.

{\bf Set 4 - J1256+3852:} This radio-loud source 
was monitored for five epochs between April 2013 and May 2014. 
The steadiest pair of stars are S2 and S3 (standard deviation
$\sim$0.0053) of which S2 is taken as the reference star. S1 is found to 
vary on the first three nights and hence is not used in the analysis for 
those nights. However, on the remaining two nights (2 February 2014 and 
1 May 2014) it is included as the second comparison star. This NLSy1 galaxy 
shows clear evidence of INOV on two nights, 5 January 2014 and 1 
February 2014, with amplitude of variability of 7.1\% and 4.8\% 
respectively (Fig.~\ref{fig:fig8}). Over a period of five months between 
5 January and 01 May, 2014, the object has steadily brightened by about 
0.075 mag.

{\bf Set 4 - J0037-0933:} Classified as radio-quiet, this source was
observed for two epochs, separated by a year. On both the nights
it is NV. For characterizing the variability, the reference star 
used is S2 and stars S1, S3 are taken as the comparison stars
(Fig.~\ref{fig:fig9}).

\begin{table}
	\centering
	\caption{Duty cycles, mean amplitude of variability and the corresponding error in amplitude (in\%), for different classes of NLSy1 galaxies}
	\label{tab:eg_table5}
	\begin{tabular}{llllll} 
		\hline
		 & \multicolumn{5}{c}{NLSy1}                 \\
                       &   GL      & GQ         & RQ       & HP      & LP \\
		(1)    & (2)    &      (3)       & (4)        & (5)      & (6)             \\
		\hline
		DC      & 55.2  & 38.8  & 0.0   &  69.0    &   58.9    \\ 
		\hline
		Mean Amp & 24.0  & 5.2   & 0.0   &    29.2   & 10.0  \\
	 		 & $\pm$13.7  & $\pm$1.6   & $\pm$0.0   &    $\pm$12.4  & $\pm$3.8 \\
		\hline
	\end{tabular}
\end{table}

\section{Results and Discussions}
Our results of INOV show clear contrast between RQ and RL-NLSy1 galaxies that includes
GL-NLSy1 and GQ-NLSy1 galaxies. None of the RQ-NLSy1 
galaxies show INOV whereas three out of the four GQ-NLSy1 galaxies show INOV. 
When compared with the INOV characteristics of GL-NLSy1 galaxies, 
all the four GL-NLSy1 galaxies show INOV, the results of which are 
already available in literature. The DC of INOV is the highest in the case of 
GL-NLSy1 galaxies with a value of about 55\%. This is followed by
GQ-NLSy1 galaxies with a DC of about 39\%. Thus, the observations reported here
find clear differences between the duty cycles of INOV in different
classes of NLSy1 galaxies. Also, GL-NLSy1 galaxies show
large amplitude variations compared to the other two classes. The 
amplitude of INOV shown by GL-NLSy1 galaxies range from 
7\% to 52\% with a mean
of 24.0 $\pm$ 13.7\%. In contrast, GQ-NLSy1 galaxies show milder
INOV ranging from 3\% to 7\%  with an average of 
5.2 $\pm$ 1.6\%. 
The non-detection of INOV in the limited sample RQ-NLSy1 galaxies
studied here, need not necessarily imply the absence of INOV in this class of 
NLSy1 galaxies. A larger sample with more monitoring observations could 
result in INOV detection.

The results obtained for NLSy1 galaxies in this work are in close
agreement with that known for quasars, with blazars showing large amplitude and
high DC of variability than radio-quiet quasars that show low DC. The contrasting INOV behaviour between
RQ and RL-NLSy1 galaxies as well as between RQ and RL-quasars can naturally
be explained by relativistic beaming arguments \citep{2003ApJ...586L..25G}. However, optical polarization  
observations in the recent past on FSRQs, separated in two groups namely
low polarization FSRQs and high polarization FSRQs, clearly show that high polarization FSRQs are more
variable (high DC and amplitude) than their less polarized counterparts.
This suggest that mere presence of relativistically beamed jet is not
sufficient to explain high INOV. Alternatively, the close
linkage between optical polarization and INOV \citep{2012A&A...544A..37G} 
indicates that optical polarization is the crucial factor for causing INOV. 
This is quite on expected lines as optical polarization
is closely connected to shocks in relativistically beamed jets in blazars.

Three of the four GL-NLSy1 galaxies considered in this work, have optical polarization observations. 
1H 0323+343 has polarization value around 1\%. However it does
show a flaring state with polarization value close to 3\% \citep{2014PASJ...66..108I}. 
For SBS 0846+513 and PMN J0948+0022, polarization as high as 10\%  
\citep{2014ApJ...794...93M} and 36\% \citep{2013ApJ...775L..26I} have been known. Optical polarization measurements are not known for other sources in our sample. Dividing the GL-NLSy1 galaxies into high polarization (HP) (with P $>$ 3\%)
and low polarization (LP) NLSy1 
galaxies, we find a DC cycle of about 69\% and 59\% for HP and LP-NLSy1 
galaxies respectively. This is in close analogy with what is seen in blazars.
\citep{2012A&A...544A..37G}. Also, GL-blazars are found to show more polarization in the optical band than GQ-blazars \citep{2016MNRAS.tmp.1329A}. In the population of GL-NLSy1 galaxies,  
sources with strong optical polarization show large amplitude
INOV (29.2 $\pm$ 12.4\%) compared to their less polarized counterparts (10.0 $\pm$ 3.8\%). However, the caveat
here is that the polarization data used here (that are compiled from literature)
has not been acquired simultaneous to our INOV observation. In spite
of this short coming, results found here clearly suggest that 
high polarization GL-NLSy1 galaxies are more variable within a night
than their counterparts with low optical polarization. To firmly establish
this point, simultaneous optical flux and polarization observations on a large number of sources are needed.
This is not feasible at present as there are only  about half a dozen 
of $\gamma$-ray emitting NLSy1 galaxies known. However, it is expected to be achievable 
in the future when more number of $\gamma$-ray emitting NLSy1 galaxies
would be detected.

In this work we want to quantify the INOV characteristics of the different categories
of NLSy1 galaxies to understand the similarities and differences among them
as well as to compare with that of their luminous counterparts, the quasars.
This is motivated by the detection of $\gamma$-ray emission from
few RL-NLSy1 galaxies by {\it Fermi}. Available studies do indicate that RL-AGN 
are hosted by black holes having high masses in excess of 
10$^8$ $M_{\odot}$ \citep{2000ApJ...543L.111L,2011MNRAS.416..917C}. The detection of $\gamma$-ray  emission in RL-NLSy1 
galaxies believed to have BH masses of the order of $10^6$ - $10^7$ $M_{\odot}$ \citep{2012AJ....143...83X}
is a direct challenge to this idea. Irrespective of this, GL-NLSy1 galaxies have now 
emerged as a new class of $\gamma$-ray emitting AGN, though the number of objects 
known as of now are limited. This is likely due the small number of 
NLSy1 galaxies known till date \citep{2006ApJS..166..128Z}. 
A multitude of observations on most of the GL-NLSy1 galaxies known
as of now, do indicate that most of the properties of these sources are 
analogous to blazars, such as large INOV \citep{2010ApJ...715L.113L,2013MNRAS.428.2450P,2016ApJ...819..121P} 
broad band SED \citep{2016ApJ...819..121P,2014ApJ...789..143P} compact core-jet structure in the 
radio \citep{2012MNRAS.426..317D,2012arXiv1205.0402O,2013MNRAS.433..952D,2014MNRAS.438.3521D}, 
super-luminal motion \citep{2013MNRAS.436..191D}, optical polarization
\citep{2014PASJ...66..108I,2013ApJ...775L..26I,2014ApJ...794...93M}  etc. All these properties do 
indicate the presence of relativistic jets in these sources, similar to blazars. As 
the broad band SED and gamma-ray spectral properties of these sources
have close resemblance to FSRQs, it has been postulated that they are the low mass black hole counterparts to FSRQs \citep{2016ApJ...819..121P}. However, this
assertion is unlikely to hold long if it could be conclusively established
that these sources do have BH masses like blazars. Recent reports do indicate
little differences between the black hole masses of GL-NLSy1 galaxies and 
blazars. Examples are the determination of BH mass of $6 \times 10^8 M_{\odot}$ in the GL-NLSy1 galaxy
PKS 2004-0447 from spectro-polarimetry \citep{2016MNRAS.458L..69B} using the
broad $H_{\alpha}$ emission line (FWHM = 9000 km/sec) in polarized light,  and 
masses of $1.6 \times 10^9 M_{\odot}$ and $3.2 \times 10^8 M_{\odot}$ respectively in PMN J0948+0022 and PKS 1502+036 via accretion disk 
modeling \citep{2013MNRAS.431..210C}. Thus, recent observations including the
one presented here point to close similarity  between GL-NLSy1 galaxies and 
blazars. However, ambiguity still remains on the host galaxy of 
GL-NLSy1 galaxies (spirals v/s ellipticals) compared to blazars that are hosted by
ellipticals. \citep{2014ApJ...795...58L,2016arXiv160902417K}

\section{Summary}
We have reported results of a first comparative study on the INOV behaviour of 
three different classes of NLSy1 galaxies, namely, RQ, GQ and GL-NLSy1 galaxies. Though the observations of RQ and GQ-NLSy1 
galaxies are reported for the first time, INOV studies of GL-NLSy1 galaxies
were available in the literature \citep{2010ApJ...715L.113L, 2013MNRAS.428.2450P,
2016ApJ...819..121P}. Taking the observations of GL-NLSy1 galaxies reported by
us earlier, the mean observing time for the sample of GL-NLSy1 galaxies
is 3.8 hours over 18 nights. For RQ-NLSy1 and GQ-NLSy1 galaxies sample
the mean monitoring time are 4.6 hours over 7 nights and 4.4 hours over 12 nights
respectively. Considering the error in the photometry 
($\sigma$ = 0.0054), 
our observations are sensitive to detect INOV above the 
1\% level corresponding to 95\% confidence (2$\sigma$), which 
forms the 
detection threshold of our observations reported here. 
The power-enhanced F-test was adopted to detect INOV. The findings of 
this study are:
\begin{enumerate}
\item INOV has been detected in RL-NLSy1 galaxies but not in RQ-NLSy1 
galaxies.
\item RL-NLSy1 galaxies that include both GL and GQ-NLSy1 galaxies 
show large DC of INOV of 50\% in sharp contrast to 
the null DC shown by RQ-NLSy1 galaxies. Thus there is a marked difference in the INOV properties between RL and RQ-NLSy1 galaxies.
This is also in close agreement with what is known of the INOV properties of
the luminous counterparts to NLSy1 galaxies, i.e. RL-quasars show INOV
more frequently than RQ-quasars. 
\item Among RL-NLSy1 galaxies, GL-NLSy1 show the highest DC of 
variability of about 55.2\% using the adopted variability statistics. Their amplitude of variability is also larger 
than the other classes of NLSy1 galaxies. They show a large mean variability 
amplitude of 24.0 $\pm$ 13.7\% compared to GQ-NLSy1 galaxies 
(5.2 $\pm$ 1.6\%).
Such extreme INOV behaviour shown by GL-NLSy1 galaxies
might be due to the high bulk Lorentz factor of their inner jets. The 
null
DC of INOV shown by RQ-NLSy1 galaxies could be due to a less powerful or a misaligned (and consequently a
lower $\delta$) jet compared to the other classes of NLSy1 galaxies.
\item Dividing the GL-NLSy1 galaxies into HP-NLSy1 and LP-NLSy1 galaxies based 
on their optical polarization properties, we find
that HP-NLSy1 galaxies show DC (amplitude) of variations of about 
69.0(29.2 $\pm$ 12.4)\%, which appear to be somewhat higher than the values of 
58.9(10.0 $\pm$ 3.8)\% shown by the LP-NLSy1 galaxies.  
\end{enumerate}

\section*{Acknowledgements}
We thank the referee Prof. Paul Wiita for his critical comments on 
the manuscript. We also thank the staff of IAO, Hanle and CREST, Hoskote, who 
made these observations possible. The facilities at IAO and CREST are operated 
by the Indian Institute of Astrophysics, Bangalore.




\bibliographystyle{mnras}
\bibliography{Master}

\bsp	
\label{lastpage}
\end{document}